\documentclass[conference]{ieeeconf}
\IEEEoverridecommandlockouts
\usepackage{cite}
\usepackage{amsmath,amssymb,amsfonts}
\usepackage{graphicx}
\usepackage{textcomp}
\usepackage[dvipsnames]{xcolor}
\def\BibTeX{{\rm B\kern-.05em{\sc i\kern-.025em b}\kern-.08em
    T\kern-.1667em\lower.7ex\hbox{E}\kern-.125emX}}

\usepackage{mleftright}

\usepackage{import}
\usepackage{algorithm}
\usepackage{algpseudocode}

\usepackage{pgfplots}
  \usepgfplotslibrary{groupplots}
  \usepgfplotslibrary{fillbetween} 
\pgfplotsset{compat=newest}

\usepackage{subcaption}
\usepackage{tikz}
\usetikzlibrary{positioning, arrows, calc, shadings}

\tikzstyle{blockdiag}	= [node distance=5mm, >=stealth', semithick]
\tikzstyle{block}			= [draw, rectangle, minimum width=1cm, minimum 
height=.8cm]
\tikzstyle{sum} = [draw,circle,inner sep=0pt, minimum size=6pt]
\tikzstyle{connector} = [draw,circle,inner sep=0pt, minimum size=2pt, 
fill=black]

\usepackage{array}% in the preamble
\newcommand{\sys}[1]{\mathbf{#1}} % XXX JT: SET HERE APPEARANCE OF OPERATORS
\newcommand{\figref}[1]{Fig.~\ref{#1}}
\newcommand{\norm}[1]{\left\|#1\right\|}

\newcommand{\field}[1]{\mathbb{#1}}
\newcommand{\R}{\field{R}}
\newcommand{\RH}{\field{RH}_\infty}

\newcommand{\Sm}{\field{S}}

\newcommand{\bmtx}{\begin{bmatrix}}
\newcommand{\emtx}{\end{bmatrix}}
\newcommand{\bsmtx}{\left[ \begin{smallmatrix}} 
\newcommand{\esmtx}{\end{smallmatrix} \right]} 
\newcommand{\bmatarray}[1]{\left[\begin{array}{#1}}
\newcommand{\ematarray}{\end{array}\right]}

\definecolor{blue1}{RGB}{222,235,247}
\definecolor{blue2}{RGB}{158,202,225}
\definecolor{blue3}{RGB}{49,130,189}

\newtheorem{mytheo}{Theorem}
\newtheorem{ex}{Example}

\newtheorem{remark}{Remark}

\newcommand{\edtn}[1]{\null} %{ {\color{red} #1} } {\null}

\begin{document}

%\title{Observer-Based Synthesis of Finite Horizon Linear Time-Varying Controllers

%AuthorBlock
%\author{\IEEEauthorblockN{1\textsuperscript{st} Felix Biertümpfel}
%\IEEEauthorblockA{\textit{Chair of Flight Mechanics and Control} \\
%\textit{Technische Universität Dresden}\\
%Dresden, Germany \\
%felix.biertuempfel@tu-dresden.de}
%\and
%\IEEEauthorblockN{2\textsuperscript{nd} Julian Theis}
%\IEEEauthorblockA{\textit{Underwater Vehicles Engineering Dept.} \\
%\textit{\textsc{Atlas Elektronik} GmbH}\\
%Bremen, Germany \\
%julian.theis@atlas-elektronik.com}
%\and
%\IEEEauthorblockN{3\textsuperscript{rd} Harald Pfifer}
%\IEEEauthorblockA{\textit{Chair of Flight Mechanics and Control} \\
%\textit{Technische Universität Dresden}\\
%Dresden, Germany \\
%harald.pfifer@tu-dresden.de}

\title{\LARGE \bf
Control Synthesis Along Uncertain Trajectories Using Integral Quadratic Constraints*
}

\author{Felix Biertümpfel$^{1}$, Peter Seiler$^{2}$ and Harald Pfifer$^{1}$% <-this % stops a space
\thanks{*This work was supported by the European Union under Grant No. 101153910. The responsibility for the content of this paper is
with its authors.}% <-this % stops a space
\thanks{$^{1}$ Felix Biertümpfel and Harald Pfifer are with the Chair of Flight Mechanics and Control, Technische Universität Dresden,
         01069 Dresden, Germany
        {\tt\small felix.biertümpfel@tu-dresden.de} {\tt\small harald.pfifer@tu-dresden.de}}%
\thanks{$^{2}$ Peter Seiler is with the Department of Electrical and Computer Engineering at the University of Michigan, Ann Arbor, {\tt\small pseiler@umich.edu}}%
}

%\and
%\and
%\and

\maketitle

\begin{abstract}
The paper presents a novel approach to synthesize robust controllers for nonlinear systems along perturbed trajectories.
The approach linearizes the system with respect to a reference trajectory. In contrast to existing methods rooted in robust linear time-varying synthesis, the approach accurately includes perturbations that drive the system away from the reference trajectory.  Hence, the controller obtained in the linear framework provides a significantly more robust nonlinear performance. The calculation of the controller is derived from robust synthesis approaches rooted in the integral quadratic constraints framework. The feasibility of the approach is demonstrated on a pitch tracker design for a space launcher. 
\end{abstract}

%\begin{IEEEkeywords}
%Robust control, Time-varying systems, Uncertain systems
%\end{IEEEkeywords}

% =================================================================================================
% ============================================= Introduction ============================================
% =================================================================================================
\vspace{-1pt}
\section{Introduction}\label{sec:Intro}
\vspace{-2pt}
Following a predefined trajectory is a fundamental engineering problem, e.g. for robotic manipulators \cite{Hosovsky2016}, space launch vehicles \cite{Biertuempfel2021a}, and planetary landers \cite{Capolupo2024}. Uncertainties and disturbances can cause the system to diverge from the predefined trajectory and degrade their performance. For instance, the ascent trajectory of a space launcher is determined for a certain wind profile and will differ substantially for any variation in the wind. Thus, a controller must provide robustness against omnipresent trajectory perturbations.

%Directly designing a controller for such a system is difficult and requires accurate models. It is also difficult to certify the performance and robustness using metrics commonly used in industry, e.g. gain and phase margins. An alternative is to rely on linear system theory.

System dynamics are usually modeled by nonlinear differential equations. The nonlinear differential equations can be linearized with respect to a reference trajectory yielding a finite horizon linear time-varying (LTV) system. Its system matrices are known functions of time and resemble the original system dynamics. There are various approaches in the literature to synthesize controllers for LTV systems. These include methods for nominal synthesis \cite{Limebeer1992, Biertumpfel2022} and robust synthesis for systems with uncertainty \cite{ OBrien1999, Pirie2002, Farhood2008}. Most recent synthesis approaches for robust controllers reside inside the integral quadratic constraint (IQC) framework~\cite{Buch2021}, which is briefly presented in Section~\ref{sec:pre}.
However, the synthesis approaches so far only consider uncertainties in the LTV state space representation. They thus fail to recover the dynamics of the nonlinear system if parameters are perturbed along the trajectory.

% For many engineering problems, uncertain parameter-variations are the main reason for departures from the reference trajectory.
%A space launcher, for example, is subject to a nominal wind profile during its atmospheric flight usually estimated shortly before flight. However, the actually encountered wind profile will differ from the nominal one due to inaccuracies in the estimations. The result is a departure from the desired state trajectory.

The present paper extends the LTV robust synthesis framework in \cite{Buch2021} to robust controllers for nonlinear systems in the vicinity of a reference trajectory. To do so, the state space representation of the LTV system is extended with a constant driving term and terms associated with the parameter variations in Section \ref{sec:Problem}. Specifically, the synthesis problem is shown in Section \ref{sec:PertTraj} to become a LTV IQC synthesis under uncertain initial conditions. Furthermore, the effects of the linearization error along the trajectory can be included in a similar way, see, e.g., \cite{Biertuempfel2023a}. Thus, a robust controller for nonlinear systems along uncertain trajectories can be calculated using linear methods. The approach is demonstrated on 
a pitch tracker control design of space launcher in Section~\ref{sec:NumEx}.
%
%In other words, an actually synthesized  bound on the worst-case performance of the nonlinear system is obtained using linear analysis tools. 
%?Similar approaches for the analysis most closely related ... IFAC, Geng, Schweidel?

%In summary, the present paper leverages existing \textcolor{blue}{LTV IQC analysis} results for worst-case performance analysis of nonlinear systems. It considers the effects of parameter variation on the state trajectory and also includes the uncertaintiy associated with the linearization error. The approach is demonstrated on a numerical example in Section~\ref{sec:NumEx}.
%\textcolor{blue}{xxxFB: das mit leverage kling super weak, wir machen ja schon was cleveres.}

%In summary, the present paper explicitly considers the effects of parameter variation on the state trajectory.
%It extends the LTV synthesis frame to robust controllers for nonlinear systems in the vicinity of a reference trajectory. The approach is demonstrated on 
%a pitch tracker control design of space launcher of industrial complexity in section~\ref{sec:NumEx}.
%
%\begin{itemize}
%	\item we do not look at uncertain initial states, but an uncertain pseudo state $x_\rho$ which acts as a forcing term 
%	\item ``nominal'' synthesis step is performed with uncertain initial pseudo state $x_\rho$, which represents a constant forcing term
%	\item  analysis is performed for uncertain initial pseudo state $x_\rho$ and a time-varying paramtric uncertainty $\rho$ which leads to a time-varying forcing term
%\end{itemize}
%

\vspace{-1.0pt}
\section{Preliminaries}\label{sec:pre}
\vspace{-1.0pt}
\subsection{Uncertain Linear Time-Varying Systems}
\vspace{-1.0pt}
%Consider an LTV system $H$ defined on the horizon $[0,T]$:
%\begin{align}
%	\label{eq:LTV}
%	\begin{split} 
%		\dot{x}(t) &= A(t)\, x(t) + B(t)\, d(t)\\
%		e(t) & = C(t) \, x(t) + D(t) \, d(t)
%	\end{split}
%\end{align}
%where $x \in \R^{n_x}$ is a state, $d \in \R^{n_d}$ is the exogenous input, and
%$e \in \R^{n_e}$ is the output.  The state matrices
%$A : [0,T] \rightarrow \R^{n_x \times n_x}$, etc. are piecewise-continuous
%(bounded) matrix valued functions of time. It is assumed throughout
%that $T < \infty$. Thus $d \in \mathcal{L}_2[0,T]$ implies $x$ and $e$
%are in $\mathcal{L}_2[0,T]$ for any initial condition $x(0)$
%\cite{brockett2015finite}.  Explicit time dependence of the state
%matrices is omitted when clear from the context.
%
%\
%
%\textcolor{blue}{nominal performance}
%
%\
%
%The performance of an LTV system $H$ can be specified in terms of its induced $L_2[0, T]$ gain \textcolor{red}{FB:better already with non-zero initial conditions?}:
% \begin{equation}
%\label{eq:finiteL2gain}
%\norm{H}_{2[0, T]} := \sup_{\substack{d \ne 0, d \in L_2[0, T], x(0)=0}} \frac{\norm{e}_{2[0,T]}}{\norm{d}_{2[0,T]}},
%\end{equation}
%with $\norm{d}_{2[0,T]} 0 [ \int_0^T d(t)^Td(t)\,dt ]^\frac{1}{2}$.
%This norm can be readily calculated following the approach in \cite{Green1995}
%
%\
%
%\textcolor{blue}{uncertain LTV system and IQC}
%
%\
%
%\textcolor{blue}{write it as short as the CEAS}

\begin{figure}[h!]
\vspace{-9.0pt}
  \centering
  % Define Colors
\definecolor{mycolor1}{HTML}{367D7D}
\definecolor{mycolor2}{HTML}{D33502}
\definecolor{mycolor3}{HTML}{FAA818}
\definecolor{mycolor4}{HTML}{41A30D}
\definecolor{mycolor5}{HTML}{FFCE38}
\definecolor{mycolor6}{HTML}{6EBCBC}
\definecolor{mycolor7}{HTML}{37526D}

\colorlet{plant}{mycolor5!40!white}
%\colorlet{unc}{mycolor6!40!white}
\colorlet{unc}{white}
%\colorlet{closedloop}{gray!20!white}
\colorlet{closedloop}{white}
\colorlet{controller}{mycolor4!30!white}
\colorlet{dist}{mycolor3!60!white}
\colorlet{marker}{mycolor7}
\colorlet{algo}{mycolor3!60!white}
\colorlet{miscplot}{mycolor2}
\colorlet{clplot}{mycolor2}

% Arrow Style
\tikzset{>=stealth'}

\begin{tikzpicture}[thick,scale=0.9,rounded corners = 0.0mm]
\draw [fill=closedloop] (0,0) rectangle node{$N$}(1.2,1.2);
\draw [fill=unc] (0.2,1.6) rectangle node{$\Delta$}(1,2.4);
\draw [->](0,1) -- (-0.5,1) -- (-0.5,2) -- (0.2,2);
\draw [->](1,2) -- (1.7,2) -- (1.7,1) -- (1.2,1);
\draw [->](0,0.5) -- (-1.2,0.5);
%\draw [->](0,0.3) -- (-1.2,0.3);
\draw [->](2.4,0.5) -- (1.2,0.5);
\node at (2.6,0.5) {$d$};
\node at (1.9,1.5) {$w$};
\node at (-0.7,1.5) {$v$};
\node at (-1.5,0.7) {$e$};
%\node at (-1.5,0.3) {$e_E$};
\end{tikzpicture}		
  \caption{Interconnection LTV system $N$ and uncertainty $\Delta$}
  \label{fig:UncertainOL}
\vspace{-12.5pt}
\end{figure}
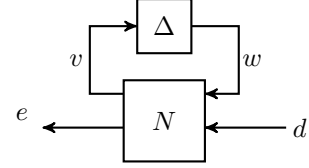

An uncertain linear time-varying system $\mathcal{F}_u(N, \Delta)$ is defined by the feedback interconnection of a nominal LTV system $N$ and the perturbation $\Delta$, also referred to as ``uncertainty'' for simplicity,  as pictured in Fig. \ref{fig:UncertainOL}. 

The known LTV system $N$ is given by:
\begin{align}
\label{eq:uLTV}
\bmtx \dot{x}(t)\\v(t)\\ e(t)\emtx =
\bmtx A(t) & B_w(t) & B_d(t)\\
C_v(t) & D_{vw}(t) & D_{vd}(t)\\
C_{e}(t) & D_{ew}(t) & D_{ed}(t)
\emtx \bmtx x(t) \\ w(t)\\ d(t)\emtx
\end{align}
%\textcolor{blue}{FB: Write with the measured output $y$ and the input $u$? This is the most general form. Thus, we can always just say ... like N but w/o certain in-/outputs}.
In (\ref{eq:uLTV}), $x(t) \in \R^{n_{x}}$, $d(t) \in \R^{n_d}$, and $e(t) \in \R^{n_e}$ denote the state, input, and the output vector, respectively. The vectors $w(t)$ and $v(t)$ connect $N$ with $\Delta$, i.e.,  $w=\Delta(v)$. The state space matrices are piecewise continuous locally bounded matrix-valued functions of time with corresponding dimensions. To shorten the notation, the explicit time dependence is omitted, if clear from the context. The uncertainty $\Delta: L_2^{n_v}[0,T] \rightarrow L_2^{n_w}[0,T]$ is a bounded and causal operator.
The interconnection $\mathcal{F}_u (N,\Delta)$ in Fig.~\ref{fig:UncertainOL} is said to be well-posed if, for all initial conditions, $x(0)$ and $d\in L_2[0,T]$ unique solutions $x\in L_2[0,T]$, $v\in L_2[0,T]$, and $w\in L_2[0,T]$ satisfying \eqref{eq:uLTV} and causally dependent on $d$ exist.
%This operator can describe nonlinearities like saturations, infinite-dimensional operators like time delays, and dynamic and real parametric uncertainties. 

Let $\mathcal{S}$ denote a set of uncertainties to be considered.  The
robust performance of the uncertain system $\mathcal{F}_u(N,\Delta)$ for uncertain initial conditions and a given $\Delta$
is quantified by
\begin{equation}
\label{eq:E2PWC}
\begin{gathered}
\!\!  \| \mathcal{F}_u(N,\Delta) \|_{2[0,T]}\!\! :=\!\!\!\! \sup_{\substack{d \in
      L_2[0,T]\\ d \neq 0, x(0) \in \R^{n_{x}}}} \!\!\!\bigg[\frac{ \norm{e}^2_{2[0,T]}}{\norm{d}^2_{2[0,T]} + \norm{x(0)}_2^2}\bigg]^2\!\!,
%\| F_u(G_t,\Delta) \|_{[0, T]} := \sup_{\substack{\Delta \in IQC(\Psi, M)}} \norm{F_u(G_t, \Delta)}_{2[0, T]}
\end{gathered}
\end{equation}
where $\norm{d}_{2[0,T]} := \left[ \int_0^T d(t)^T d(t) \, dt \right]^{\frac{1}{2}}$ is the $L_2$ norm on the finite horizon $[0,T]$.
The worst-case gain over the uncertainty set is $\sup_{ \Delta \in \mathcal{S}}\| \mathcal{F}_u(N,\Delta) \|_{2[0,T]}$.
%\begin{equation}
%\label{eq:E2PWC}
%\begin{gathered}
% \sup_{ \Delta \in \mathcal{S}}\| F_u(N,\Delta) \|_{2[0,T]}.
%\end{gathered}
%\end{equation}

%where the operator $\norm{.}$ denotes the finite horizon $L_2[0,T]$ norm  $\big[ \int_0^T d^T(t)d(t)\,dt \big]^\frac{1}{2}$. %\textcolor{red}{FB: directly with $x(0)^TRx(0)$ and $x(0)=x_0 \in \R^{n_x}$?}
%\begin{equation}
%\label{eq:E2PWC}
%\begin{gathered}
%  \| F_u(N,\Delta) \|_{2[0,T]}\! := \sup_{ \Delta \in \mathcal{S}}\!\!\! \sup_{\substack{d \in
%      L_2[0,T]\\ d \neq 0, x(0) \in \R^{n_{x}}}} \!\!\!\bigg[\frac{ \norm{e}^2_{2[0,T]}}{\norm{d}^2_{2[0,T]} + \norm{x(0)}_2^2}\bigg]^2\!,
%%\| F_u(G_t,\Delta) \|_{[0, T]} := \sup_{\substack{\Delta \in IQC(\Psi, M)}} \norm{F_u(G_t, \Delta)}_{2[0, T]}
%\end{gathered}
%\end{equation}

The input/output behavior of $\Delta$ is bounded via time-domain IQCs.
Time-domain IQCs are defined by a filter $\Psi \in \RH^{n_z \times (n_v + n_w)}$ with input $[v^T, w^T]^T$ and output $z$, and a $n_z \times n_z$ real, symmetric matrix $M$, see, e.g., \cite{Seiler2015}. 
%The filter $\Psi$ and the matrix $M$ result from a (non-unique) factorization  $\Pi = \Psi^\sim M \Psi$ \cite{Scherer2004} of the classical frequency-domain IQC multipliers $\Pi$ proposed by Megretski and Rantzer \cite{Megretski1997}. Hence, $\Psi$ can be interpreted as a $D$-scale common in robust control literature\cite{Packard1993}.
The uncertainty $\Delta$ satisfies the time domain IQC defined by $M$ and $\Psi$ if the filter output $z$ fulfills $\int_0^T z(t)^T M z(t) \, dt \ge 0 $
%\begin{equation}
%    \label{eq:iqctd1}
%    \int_0^T z(t)^T M z(t) \, dt \ge 0
%\end{equation}
for all $v \in L_2[0,T]$, $w=\Delta(v)$ over the interval $[0,T]$ for zero initial conditions.
In this case, the short notation $\Delta \in IQC(\Psi,M)$ is used.
An example for feasible parameterization and factorization of a time-varying parametric uncertainty is given below.
\begin{ex}
  \label{ex1}
  Let $\mathcal{S}$ denote the set of nonlinear, time-varying,
  uncertainties with a given norm-bound $b$, i.e.
  $\Delta \in \mathcal{S}$ if
  $ \|\Delta\|_{2\rightarrow 2,[0,T]} \leq b$. If
  $\Delta \in \mathcal{S}$, then for any constant $\lambda\ge 0$ it
  satisfies the IQC defined by the static filter $\Psi =I_{n_v + n_w}$ and constant matrix:
  \begin{align*}
    M := \bsmtx b^2 \lambda \, I_{n_v} & 0\\0 & -\lambda \, I_{n_w}\esmtx	
  \end{align*}
\end{ex}

\

\vspace{-6.5pt}
\subsection{Robust Synthesis}\label{ss:RobSyn}
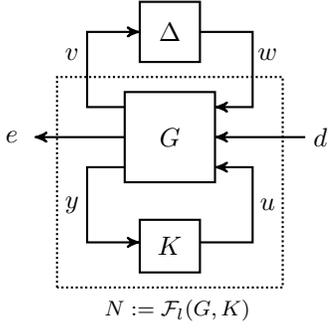
\begin{figure}
  \centering
  % Define Colors
\definecolor{mycolor1}{HTML}{367D7D}
\definecolor{mycolor2}{HTML}{D33502}
\definecolor{mycolor3}{HTML}{FAA818}
\definecolor{mycolor4}{HTML}{41A30D}
\definecolor{mycolor5}{HTML}{FFCE38}
\definecolor{mycolor6}{HTML}{6EBCBC}
\definecolor{mycolor7}{HTML}{37526D}

%\colorlet{plant}{mycolor5!40!white}
\colorlet{plant}{white}
%\colorlet{unc}{mycolor6!40!white}
\colorlet{unc}{white}
%\colorlet{closedloop}{gray!20!white}
\colorlet{closedloop}{white}
%\colorlet{controller}{mycolor4!30!white}
\colorlet{controller}{white}
\colorlet{dist}{mycolor3!60!white}
\colorlet{marker}{mycolor7}
\colorlet{algo}{mycolor3!60!white}
\colorlet{miscplot}{mycolor2}
\colorlet{clplot}{mycolor2}

% Arrow Style
\tikzset{>=stealth'}

\begin{tikzpicture}[thick,rounded corners = 0.0mm]
\begin{scope}[shift={(-0.2,0)}]
\draw[densely dotted,fill=closedloop,fill opacity=1] (-0.7,-1.4) rectangle (2.3,1.4);
\end{scope}
\draw [fill=plant] (0,0) rectangle node{$G$}(1.2,1.2);
\draw [fill=unc] (0.2,1.6) rectangle node{$\Delta$}(1,2.4);
\draw [->](0,1) -- (-0.5,1) -- (-0.5,2) -- (0.2,2);
\draw [->](1,2) -- (1.7,2) -- (1.7,1) -- (1.2,1);
\draw [->](0,0.6) -- (-1.2,0.6);
%\draw [->](0,0.4) -- (-1.2,0.4);
\draw [->](2.4,0.6) -- (1.2,0.6);
\node at (2.6,0.6) {$d$};
\node at (1.9,1.7) {$w$};
\node at (-0.7,1.7) {$v$};
\node at (-1.5,0.6) {$e$};
%\node at (-1.5,0.4) {$e_E$};
\draw [fill=controller] (0.2,-0.5) rectangle node{$K$}(1,-1.2);
\draw [->](0,0.2) -- (-0.5,0.2) -- (-0.5,-0.8) -- (0.2,-0.8);
\draw [->](1,-0.8) -- (1.7,-0.8) -- (1.7,0.2) -- (1.2,0.2);
\node at (-0.7,-0.3) {$y$};
\node at (1.9,-0.3) {$u$};
\node at (0.7,-1.7) {\footnotesize$N:=\mathcal{F}_l(G,K)$};
\end{tikzpicture}		
  \caption{Uncertain closed loop $\mathcal{F}_u(\mathcal{F}_l(G,K),\Delta)$}
  \label{fig:UncertainCL}
\vspace{-20.0pt}
\end{figure}

\vspace{-2.5pt}
\figref{fig:UncertainCL} shows the feedback interconnection $\mathcal{F}_u(\mathcal{F}_l(G,K),\Delta)$ of an uncertain LTV system and the controller $K$, with $N:=\mathcal{F}_l(G,K)$ corresponding to~\eqref{eq:uLTV}.
%The known part of the closed loop system, i.e., $\mathcal{F}_l(G,K)$ corresponds to~\eqref{eq:uLTV}.
%The known part of the system $G$ has the same form as~\eqref{eq:uLTV}, but with an additional output $y$ (measured outputs) and input $u$ (inputs from controller).
The finite horizon robust synthesis problem
is to find a linear time-varying controller simultaneously minimizing the impact of
worst-case disturbances and worst-case uncertainties, i.e.:
\begin{align}
  \label{eq:robsynOrg}
  \inf_{K} \sup_{\Delta \in \mathcal{S}} 
  \|\mathcal{F}_u(\mathcal{F}_l(G,K),\Delta)\|_{2[0,T]}
\end{align}
%Specifically, the goal is to design a linear
%time-varying controller
%$K:\mathcal{L}^{n_y}_2[0,T]\rightarrow\mathcal{L}^{n_u}_2[0,T]$ such
%that the worst-case gain is bounded by
%$\gamma$, i.e.:
%\begin{align}
%  \label{eq:robBnd1}
%  \|\mathcal{F}_u(\mathcal{F}_l(G,K),\Delta)\|_{2[0,T]} < \gamma
%  \,\,\, \forall \Delta \in \mathcal{S}
%\end{align}

In general, problem~\eqref{eq:robsynOrg} is non-convex. Reference \cite{Buch2021} provides an iterative approach to synthesize 
a robust LTV controller for zero initial conditions. Particularly, the procedures iterates through three essential steps: 1) Construction of the scaled plant using the IQC data $(\Psi, M)$, 2) Nominal
LTV synthesis using the scaled plant, and 3) IQC-based robust performance analysis of the uncertain closed loop $\mathcal{F}_u(\mathcal{F}_l(G,K),\Delta)$.
The steps are briefly presented in the following and the reader is referred to \cite{Buch2021} for more detailed explanations.

\
\vspace{-9.0pt}
\subsubsection{Scaled Plant Construction}
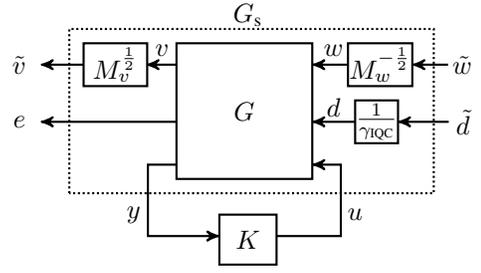
\begin{figure}
  \centering
  % Define Colors
\definecolor{mycolor1}{HTML}{367D7D}
\definecolor{mycolor2}{HTML}{D33502}
\definecolor{mycolor3}{HTML}{FAA818}
\definecolor{mycolor4}{HTML}{41A30D}
\definecolor{mycolor5}{HTML}{FFCE38}
\definecolor{mycolor6}{HTML}{6EBCBC}
\definecolor{mycolor7}{HTML}{37526D}

\colorlet{plant}{mycolor5!40!white}
\colorlet{unc}{mycolor6!40!white}
\colorlet{closedloop}{gray!20!white}
\colorlet{controller}{mycolor4!30!white}
\colorlet{dist}{mycolor3!60!white}
\colorlet{marker}{mycolor7}
\colorlet{algo}{mycolor3!60!white}
\colorlet{miscplot}{mycolor2}
\colorlet{clplot}{mycolor2}

% Arrow Style
\tikzset{>=stealth'}

\begin{tikzpicture}[thick,scale=0.95,rounded corners = 0.0mm]
\begin{scope}[shift={(-0.2,0)}]
%\draw[densely dotted,fill=blue,fill opacity=0.1] (-1.3,-0.9) rectangle (3.8,1.4);
\draw[densely dotted] (-1.3,-0.9) rectangle (3.8,1.4);
\end{scope}
%\draw [fill=plant] (0,-0.7) rectangle node{$G$}(1.9,1.2);
%\draw [fill=dist] (2.5,-0.2) rectangle node{$\frac{1}{\gamma_a}$}(3.1,0.4);
%\draw [fill=unc] (2.4,0.6) rectangle node{$M_{w}^{-\frac{1}{2}}$}(3.3,1.2);
%\draw [fill=unc] (-1.3,0.6) rectangle node{$M_{v}^{\frac{1}{2}}$}(-0.41,1.2);
\draw (0,-0.7) rectangle node{$G$}(1.9,1.2);
\draw (2.5,-0.2) rectangle node{$\frac{1}{\gamma_\text{IQC}}$}(3.1,0.4);
\draw (2.4,0.6) rectangle node{$M_{w}^{-\frac{1}{2}}$}(3.3,1.2);
\draw (-1.3,0.6) rectangle node{$M_{v}^{\frac{1}{2}}$}(-0.41,1.2);
\draw [->](0,0.1) -- (-1.9,0.1);
%\draw [->](0,0) -- (-1.9,0);
\draw [->](0,0.9) -- (-0.42,0.9);
\draw [->](2.5,0.1) -- (1.9,0.1);
\draw [->](2.4,0.9) -- (1.9,0.9);
\draw [->](3.8,0.9) -- (3.3,0.9);
\draw [->](3.81,0.1) -- (3.1,0.1);
\draw [->](-1.3,0.9) -- (-1.9,0.9);
\node at (2.2,0.3) {$d$};
\node at (4.01,0.1) {$\tilde{d}$};
\node at (1,1.6) {$G_\text{s}$};
\node at (2.2,1.1) {$w$};
\node at (4,0.9) {$\tilde{w}$};
\node at (-0.2,1.1) {$v$};
\node at (-2.2,0.9) {$\tilde{v}$};
\node at (-2.2,0.1) {$e$};
%\node at (-2.2,0) {$e_E$};
%\draw [fill=controller] (0.6,-1.2) rectangle node{$K$}(1.4,-1.9);
\draw (0.6,-1.2) rectangle node{$K$}(1.4,-1.9);
\draw [->](0,-0.5) -- (-0.4,-0.5) -- (-0.4,-1.5) -- (0.6,-1.5);
\draw [->](1.4,-1.5) -- (2.3,-1.5) -- (2.3,-0.5) -- (1.9,-0.5);
\node at (2.5,-1.2) {$u$};
\node at (-0.6,-1.2) {$y$};
\end{tikzpicture}			
  \caption{Scaled Plant $G_\text{s}$}
  \label{fig:ScaledPlant}
\vspace{-20.0pt}
\end{figure}

The scaled plant $G_\text{s}^{(i)}$ at an iteration step $i$ is constructed as shown in
\figref{fig:ScaledPlant} by scaling the performance and
uncertainty channels of $G$ using $M_{v}^{(i-1)}$,
$M_{w}^{(i-1)}$ and the worst case upper bound $\gamma_\text{IQC}^{(i-1)}$ from the previous
iteration or user-specified initial values for the first iteration. This scaling ensures appropriate normalization of the
performance and uncertainty channels. This is a key step which
connects the nominal synthesis (step 2) and worst-case gain problem (step 3). Let the
uncertain plant \eqref{eq:uLTV} be given as shown in
\figref{fig:UncertainOL}. The superscripts
$(i-1)$ are dropped in the following for better readability. The IQC matrix is assumed constant with the particular form
$M = \mbox{diag}\{M_{v},-M_{w}\}$ where $M_{v}> 0$ and
$M_{w}> 0$. Time-varying IQC-matrices are not pursued in this paper, but are covered in, e.g., \cite{Buch2021}.
Additionally, the worst-case gain
bound scales the corresponding disturbance channel $\tilde{d}=\gamma_\text{IQC}^{(i-1)}\,d$.
The scaled plant $G_\text{s}$
is constructed as shown in \figref{fig:ScaledPlant}. The scaled plant 
has nominal performance $\|G_\text{s}\|_{2[0,T]}\leq 1$.
%The worst-case
%gain upper bound DLMI \eqref{DLMI} can be re-written to demonstrate
%that this scaled plant satisfies nominal performance
%$\|N_{scl}\|_{[0,T]}\leq 1$.

\
\vspace{-9.0pt}
\subsubsection{Nominal Synthesis}
The next step is to perform finite horizon
nominal synthesis on the scaled plant.  This step is performed using
a special case of the synthesis results described in~\cite{Green1995}. This yields a controller $K^{(i)}$ minimizing the induced norm
from $[\tilde{w}^T, \tilde{d}^T]^T$ to $[\tilde{v}^T, e^T]$.

\
\vspace{-9.0pt}
\subsubsection{IQC Analysis}
Finally, the the worst-case gain upper bound 
$\gamma_\text{IQC}^{(i)}$ of the uncertain closed loop $\mathcal{F}_u(\mathcal{F}_l(G,K),\Delta)$ is computed. Note that this step
uses the original uncertain plant without any scalings.  
A differential linear matrix inequality (DLMI) can be used to calculate an upper bound on
the worst-case gain.
%\begin{align}
%  \label{DLMI}
%  \bsmtx
%    \dot{P}+A^TP+PA & PB\\
%    B^TP & 0 
%  \esmtx + \bsmtx
%    Q & S\\
%    S^T & R
%  \esmtx
%  + \bsmtx C_v^T\\ D_v^T \esmtx M \bsmtx
%  C_v^T \\ D_v^T
%	\esmtx^T \leq -\epsilon I
%\end{align}
Theorem 6 in Reference \cite{Seiler2019} states a sufficient DLMI condition to bound the worst case performance. The corresponding algorithm is used in
\cite{Buch2021} to calculate $\gamma_\text{IQC}^{(i)}$. All subsequent iterations require
the construction of a scaled plant using the IQC results and ultimately links the
analysis and synthesis step.

\vspace{-12.5pt}
\section{Problem Formulation}\label{sec:Problem}
\vspace{-10.5pt}
%
%\begin{itemize}
%	%\item Story comparable to IFAC paper
%	\item Nonlinear system along trajectory NOT LTV system
%	\item Linearization full including parameter $\rho_\Delta$
%	\item Controller is for the nonlinear system as we cover "all" effects% regler fuer das nichtlineare system da wir alle effekte it dabei haben
%	\item can be easily extended to include different types of uncertainties, non linearities, saturations or infinite dimensional systems such as time delays
%\end{itemize}
%

Consider a nonlinear system $\tilde{N}$ whose dynamics are defined by the nonlinear ordinary differential equations (ODEs)
\begin{equation}\label{eq:NonLin}
	\begin{split}
		\dot{x}(t) &= f(x(t), d(t), u(t),\rho(t)) \\
		e(t) &= h(x(t), d(t), u(t), \rho(t))\\
		y(t) &= g(x(t), d(t), u(t), \rho(t)).\\
	\end{split}
\end{equation}
% the vectors $x(t)\in \R^{n_x}$, $d(t)\in \R^{n_d}$, and $e(t)\in \R^{n_e}$ describe the state variable, the  input, and the output at time $t$, respectively. 
In (\ref{eq:NonLin}), the vector $\rho(t)\in \R^{n_\rho}$ denotes a time-varying parameter vector at time $t$. A pre-defined trajectory is defined as a unique solution of (\ref{eq:NonLin}) satisfying
\begin{equation}\label{eq:NonLinRef}
	\begin{split}
		\dot{x}_\mathcal{T}(t) &= f(x_\mathcal{T}(t), d_\mathcal{T}(t), u_\mathcal{T}(t), \rho_\mathcal{T}(t)) \\
		e_\mathcal{T}(t) &= h(x_\mathcal{T}(t), d_\mathcal{T}(t), u_\mathcal{T}(t), \rho_\mathcal{T}(t)) \\
y_\mathcal{T}(t) &= g(x_\mathcal{T}(t), d_\mathcal{T}(t), u_\mathcal{T}(t), \rho_\mathcal{T}(t)) \quad \forall t\in [0,T].
	\end{split}
\end{equation}
The subscript $\mathcal{T}$ will be used to denote a specific trajectory, $(x_\mathcal{T}, \, d_\mathcal{T}, \, u_\mathcal{T},\, \rho_\mathcal{T} )$ satisfying \eqref{eq:NonLinRef}. The objective of the paper is to design a controller $K$, which in feedback interconnection with the system $\tilde{N}$ (\ref{eq:NonLin}) satisfies given performance requirements along a prescribed trajectory over the finite time horizon $[0,T]$. The resulting closed loop system denoted by $\mathcal{F}_l(\tilde{N},K)$ shall be robust against perturbations in both the input signal $d_\Delta$ and parameter signal $\rho_\Delta$:
\begin{equation}
d_\Delta  := d -d_\mathcal{T}, \quad \rho_\Delta := \rho -\rho_\mathcal{T}.
\end{equation}
\vspace{-2.0pt}The parameter perturbation $\rho_\Delta$ is confined, at each point in time, to a set $\mathcal{P} \subseteq \R^{n_\rho}$.

Defining a robust performance certificate for the control design process of a nonlinear closed loop is difficult. 
The alternative is to synthesize a robust controller using methods for finite horizon LTV systems. This requires a linearization of \eqref{eq:NonLin} via Taylor series expansion of $f$, $h$, and $g$ about the reference trajectory $\mathcal{T}$. Define $x_\Delta(t) := x(t) -x_\mathcal{T}(t)$ and $u_\Delta(t) := u(t) -u_\mathcal{T}(t)$. The expansion is
%%\begin{equation}\label{eq:Taylor}
%%	\begin{split}
%%		\dot{x}
%%		%\dot{\bar{x}} + \dot{\tilde{x}} 
%%		=& \dot{x}_\mathcal{T} + A x_\Delta  
%%		+B_d\,d_\Delta
%%		+B_u\,u_\Delta 
%%		+E\,\rho_\Delta
%%		+ \epsilon_f
%%		\\
%%		e
%%		%\dot{\bar{e}} + \dot{\tilde{e}} 
%%		=& 	 e_\mathcal{T} +C_e\,x_\Delta  
%%		+D_{ed}\,d_\Delta 
%%		+D_{eu}\,u_\Delta 
%%		+F_e\,\rho_\Delta
%%		+ \epsilon_{h_1},
%%\\
%%y
%%		%\dot{\bar{e}} + \dot{\tilde{e}} 
%%		=& 	 y_\mathcal{T} +C_y\,x_\Delta  
%%		+D_{yd}\,d_\Delta 
%%		+D_{yu}\,u_\Delta 
%%		+F_y\,\rho_\Delta
%%		+ \epsilon_{h_2}.
%%	\end{split}
%%\end{equation}
%%Here,  $\epsilon_f$ and $\epsilon_h$ denote higher-order terms in the Taylor expansion. Moreover, the state space matrices are given by Jacobians evaluated along the $\mathcal{T}$, e.g.,  $A:=\left.\frac{\partial f}{\partial x}\right|_\mathcal{T}$. 

%%%%%%%%%%%%%%% This the whole one but is way to big to fit it in
\vspace{-13.0pt}
\begin{equation}\label{eq:Taylor}
	\begin{split}
		\dot{x}
		%\dot{\bar{x}} + \dot{\tilde{x}} 
		\!=& \dot{x}_\mathcal{T} \!+\!\!\overbrace{\left.\frac{\partial f}{\partial x}\right|_\mathcal{T}}^{A}\,x_\Delta  
		\!+\!\!\overbrace{\left.\frac{\partial f}{\partial d}\right|_\mathcal{T}}^{B_d}\,d_\Delta
		\!+\!\!\overbrace{\left.\frac{\partial f}{\partial u}\right|_\mathcal{T}}^{B_u}\,u_\Delta 
		\!+\!\!\overbrace{\left.\frac{\partial f}{\partial \rho}\right|_\mathcal{T}}^{E}\,\rho_\Delta
		\!+\! \epsilon_f
		\\
		e
		%\dot{\bar{e}} + \dot{\tilde{e}} 
		\!=& 	 e_\mathcal{T} \!+\!\!\overbrace{\left.\frac{\partial h}{\partial x}\right|_\mathcal{T}}^{C_e}\,x_\Delta  
		\!+\!\!\overbrace{\left.\frac{\partial h}{\partial d}\right|_\mathcal{T}}^{D_{ed}}\,d_\Delta 
		\!+\!\!\overbrace{\left.\frac{\partial h}{\partial d}\right|_\mathcal{T}}^{D_{eu}}\,u_\Delta 
		\!+\!\!\overbrace{\left.\frac{\partial h}{\partial \rho}\right|_\mathcal{T}}^{F_e}\,\rho_\Delta
		\!+\! \epsilon_{h},
\\
y
		%\dot{\bar{e}} + \dot{\tilde{e}} 
		\!=& 	 y_\mathcal{T} \!+\!\!\overbrace{\left.\frac{\partial g}{\partial x}\right|_\mathcal{T}}^{C_y}\,x_\Delta  
		\!+\!\!\overbrace{\left.\frac{\partial g}{\partial d}\right|_\mathcal{T}}^{D_{yd}}\,d_\Delta 
		\!+\!\!\overbrace{\left.\frac{\partial g}{\partial u}\right|_\mathcal{T}}^{D_{yu}}\,u_\Delta 
		\!+\!\!\overbrace{\left.\frac{\partial g}{\partial \rho}\right|_\mathcal{T}}^{F_y}\,\rho_\Delta
		\!+\! \epsilon_{g}.
	\end{split}
\end{equation} 
Here,  $\epsilon_f$, $\epsilon_h$, and $\epsilon_g$ denote higher-order terms in the Taylor expansion. Moreover, $\left.\frac{\partial f}{\partial x}\right|_\mathcal{T}$ denotes the Jacobian of $f$ with respect to $x$ evaluated along $\mathcal{T}$.
%%%%%%%%%%%%%
%%%%%%%%%%%%%
The common LTV state space representation neglects the higher order terms $\epsilon_f$, $\epsilon_h$, $\epsilon_g$, and assumes nominal parameters, i.e., $\rho_\Delta(t)=0$ for $t\in[0,T]$.
This approximation $\bar{N}$ is equivalent to the LTV system $N$ in \eqref{eq:uLTV} for $v = w = 0$, and additional input $u$ and output $y$.
%\begin{equation}\label{eq:StateSpaceFull}
%\begin{split}
%\dot{x}_\Delta(t) &= A(t)\,x_\Delta(t) + B(t)\,d_\Delta(t)\\
%e_\Delta(t) &= C(t)\,x_\Delta(t) + D(t)\,d_\Delta(t)
%\end{split}
%\end{equation}
%is obtained by neglecting the higher-order terms $\epsilon_f$ and $\epsilon_h$, and assuming nominal parameters, i.e., $\rho_\Delta(t)=0$ for $t\in[0,T]$.
%In equation \eqref{eq:LTV}, the perturbed output is $e_\Delta=e-e_\mathcal{T}$  and the system matrices are
%\begin{equation}
%\begin{split}
%A(t) &= \left.\frac{\partial f}{\partial x}\right|_\mathcal{T},\, B(t) = \left.\frac{\partial f}{\partial d}\right|_\mathcal{T}, \\
%C(t) &= \left.\frac{\partial h}{\partial x}\right|_\mathcal{T},\, D(t) = \left.\frac{\partial h}{\partial d}\right|_\mathcal{T}.
%\end{split}
%\end{equation}

A performance metric for $\mathcal{F}_l(\bar{N}, K)$ can be defined as 
\begin{equation} \label{eq:nomgain}
	\sup_{\substack{d_\Delta \in L_2[0,T]\\ d_\Delta \neq 0, x_\Delta(0) {=0}}}
	\frac{ \norm{e_\Delta}_{2[0,T]}}{\norm{d_\Delta}_{2[0,T]}},
\end{equation}
%\vspace{-1pt}
which is, in fact, the nominal formulation of \eqref{eq:E2PWC}.
	A controller $K$ minimizing the closed loop gain can be readily synthesized for such a system using the results in, e.g.,~\cite{Limebeer1992}.
The gain \eqref{eq:nomgain} relates perturbations in the input to perturbations from the nominal output $e_\mathcal{T}$ with respect to their $L_2[0,T]$ norm. 
%For a given norm of $d_\Delta$, a connection to \eqref{eq:wcgain} can be established, 
Since the reference output is known and $\norm{e}_{2[0,T]}\leq\norm{e_\mathcal{T}}_{2[0,T]}+\norm{e_\Delta}_{2[0,T]}$. The result can, in principle, be used to establish an upper bound on the performance of the nonlinear closed loop.
However, no actual performance guarantees are provided as this synthesis does not consider parameter perturbations and further neglects the higher-order terms of the Taylor series expansion

\begin{remark} %HP somebody with better tex skills should get rid of the number
In an attempt to cover the effects of perturbations and linearization errors, uncertainties $\Delta$ are typically added in LFT fashion to the nominal LTV system $\bar{N}$, i.e., $\mathcal{F}_u(N, \Delta)$, as seen in Section~\ref{sec:pre}. 
%Examples are dynamic uncertainties in~\cite{Seiler2019} or parametric uncertainties in~\cite{Biertuempfel2023}. 
The resulting uncertain LTV system equals \eqref{eq:uLTV}. A robust controller can be synthesized following the approach described in Section \ref{sec:pre}.
However, the signal injected into $\Delta$ remains zero as long as $d_\Delta$ is zero. Hence uncertainty modeled in this way cannot, on its own, drive states away from the nominal trajectory. This contradicts the behavior of the nonlinear system~\eqref{eq:NonLin}, where a parameter perturbation $\rho_\Delta$ is enough to drive the system away from the reference trajectory. Thus, the controller provides no robustness against trajectory perturbations.  Alternatively, the parameter variation can be treated as an external disturbance, see, e.g., \cite{Biertuempfel2021a}. However, this limits the perturbation to be part of the norm-bounded input signal $d_\Delta$.
\end{remark}

\vspace{-1.5pt}
\section{Synthesis Along Uncertain Trajectories} \label{sec:PertTraj}
\vspace{-1.5pt}
\subsection{Synthesis Problem}
\vspace{-2.0pt}
Unlike the common approach described so far, the synthesis in this paper explicitly considers perturbations from the reference trajectory, i.e.,  $\rho_\Delta(t) \neq 0$. Hence, it can be used to synthesize a controller providing robust performance of the nonlinear closed loop system. The linearization errors $\epsilon_f$, $\epsilon_h$, and $\epsilon_g$ are not explicitly considered, to make the section more lucid. However, they can be included into the synthesis following the same steps as for $\rho_\Delta$. Details are provided in \cite{Biertuempfel2023a}. % put a side 

An alternative state space representation of~\eqref{eq:Taylor} is obtained by extending the state vector $x_\Delta$ with a constant driving term:
\vspace{-3.0pt}
%%%\begin{equation}\label{eq:SSalt}
%%%\begin{split}
%%%\bsmtx \dot{x}_\Delta \\ 0 \esmtx &= \bsmtx A & E\rho_\Delta\\ 0 & 0 \esmtx \bsmtx x_\Delta \\ 1 \esmtx + \bsmtx B_d & B_u\\ 0 & 0 \esmtx \bsmtx d_\Delta \\ u_\Delta\esmtx  \\
%%%\bsmtx e_\Delta \\ y_\Delta \esmtx&= \bsmtx C_e & F_e \rho_\Delta \\ C_y & F_y \rho_\Delta \esmtx \bsmtx x_\Delta \\ 1 \esmtx+ \bsmtx D_{ed} & D_{eu}\\ D_{yd} & D_{yu}\esmtx \, \bsmtx d_\Delta \\ u_\Delta \esmtx
%%%\end{split}
%%%\end{equation}
\begin{equation}\label{eq:SSalt}
\begin{split}
\bsmtx \dot{x}_\Delta \\ 0 \esmtx &= \bsmtx A & E\rho_\Delta\\ 0 & 0 \esmtx \bsmtx x_\Delta \\ 1 \esmtx + \bsmtx B_d & B_u\\ 0 & 0 \esmtx \bsmtx d_\Delta \\ u_\Delta\esmtx  \\
\bsmtx e_\Delta \\ y_\Delta \esmtx&= \bsmtx C_e & F_e \rho_\Delta \\ C_y & F_y \rho_\Delta \esmtx \bsmtx x_\Delta \\ 1 \esmtx+ \bsmtx D_{ed} & D_{eu}\\ D_{yd} & D_{yu}\esmtx \, \bsmtx d_\Delta \\ u_\Delta \esmtx
\end{split}
\end{equation}
This model is equivalent to \eqref{eq:Taylor}. However,~\eqref{eq:SSalt} is non-standard due to the constant driving term extending the original state vector. 
This reformulation, however, enables parameter variations to excite the system dynamics even if no external disturbances occur. 
%The next step is to replace the driving term by a (scalar) pseudo-state $x_\rho$ that is confined to lie in the set $\mathcal{U} := [-1, 1]$ yielding the extended system:
The next step is to replace the driving term by a (scalar) pseudo-state $x_\rho:=1$ yielding the extended system:
%%%%%%%% no way in this form to include the (t) fully
%\begin{equation}\label{eq:SSalt2}
%\begin{split}
%\!\bmtx \dot{x}_\Delta(t) \\ 0 \emtx &\!\!=\!\! \bmtx A(t) \!\!\!&\!\!\! E(t)\rho_\Delta(t)\\ 0 \!\!\!&\!\!\! 0 \emtx \!\!\! \bmtx x_\Delta(t) \\ x_\rho \emtx \!\!+\!\! \bmtx B_d(t) \!\!\!&\!\!\! B_u(t)\\ 0 \!\!\!&\!\!\! 0 \emtx \!\!\! \bmtx d_\Delta(t) \\ u_\Delta(t)\emtx  \\
%\!\bmtx e_\Delta(t) \\ y_\Delta(t) \emtx&\!\!=\!\! \bmtx C_e(t) \!\!\!&\!\!\! F_e(t) \rho_\Delta(t) \\ C_y(t) \!\!\!&\!\!\! F_y(t) \rho_\Delta(t) \emtx \!\!\! \bmtx x_\Delta(t) \\ x_\rho \emtx\!\!\!+\!\!\! \bmtx D_{ed}(t) \!\!\!&\!\!\! D_{eu}(t)\\ D_{yd}(t) \!\!\!&\!\!\! D_{yu}(t)\emtx \!\!\! \bmtx d_\Delta(t) \\ u_\Delta(t) \emtx
%\end{split}
%\end{equation}
%%\begin{equation}\label{eq:SSalt2}
%%\begin{split}
%%\!\bmtx \dot{x}_\Delta \\ 0 \emtx &\!\!=\!\! \bmtx A(t) \!\!\!&\!\!\! E(t)\rho_\Delta(t)\\ 0 \!\!\!&\!\!\! 0 \emtx \!\!\! \bmtx x_\Delta \\ x_\rho \emtx \!\!+\!\! \bmtx B_d(t) \!\!\!&\!\!\! B_u(t)\\ 0 \!\!\!&\!\!\! 0 \emtx \!\!\! \bmtx d_\Delta \\ u_\Delta\emtx  \\
%%\!\bmtx e_\Delta \\ y_\Delta \emtx&\!\!=\!\! \bmtx C_e(t) \!\!\!&\!\!\! F_e(t) \rho_\Delta(t) \\ C_y(t) \!\!\!&\!\!\! F_y(t) \rho_\Delta(t) \emtx \!\!\! \bmtx x_\Delta \\ x_\rho \emtx\!\!\!+\!\!\! \bmtx D_{ed}(t) \!\!\!&\!\!\! D_{eu}(t)\\ D_{yd}(t) \!\!\!&\!\!\! D_{yu}(t)\emtx \!\!\! \bmtx d_\Delta \\ u_\Delta \emtx
%%\end{split}
%%\end{equation}
\vspace{-1.0pt}
%%\begin{equation}\label{eq:SSalt2}
%%\begin{split}
%%\bsmtx \dot{x}_\Delta \\ 0 \esmtx &= \bsmtx A & E\rho_\Delta\\ 0 & 0 \esmtx \!\! \bsmtx x_\Delta \\ x_\rho \esmtx \!\!+ \!\! \bsmtx B_d & B_u\\ 0 & 0 \esmtx \!\! \bsmtx d_\Delta \\ u_\Delta\esmtx  \\
%%\bsmtx e_\Delta \\ y_\Delta \esmtx&= \bsmtx C_e & F_e \rho_\Delta(t) \\ C_y & F_y \rho_\Delta \esmtx \!\! \bsmtx x_\Delta \\ x_\rho \esmtx \!\! +\!\! \bsmtx D_{ed} & D_{eu}\\ D_{yd} & D_{yu}\esmtx \!\! \bsmtx d_\Delta \\ u_\Delta \esmtx
%%\end{split}
%%\vspace{-0.0pt}
%%\end{equation}
\begin{equation}\label{eq:SSalt2}
\begin{split}
\bsmtx \dot{x}_\Delta \\ 0 \esmtx &= \bsmtx A & E\rho_\Delta\\ 0 & 0 \esmtx  \bsmtx x_\Delta \\ x_\rho \esmtx +  \bsmtx B_d & B_u\\ 0 & 0 \esmtx \bsmtx d_\Delta \\ u_\Delta\esmtx  \\
\bsmtx e_\Delta \\ y_\Delta \esmtx&= \bsmtx C_e & F_e \rho_\Delta \\ C_y & F_y \rho_\Delta \esmtx  \bsmtx x_\Delta \\ x_\rho \esmtx  + \bsmtx D_{ed} & D_{eu}\\ D_{yd} & D_{yu}\esmtx \bsmtx d_\Delta \\ u_\Delta \esmtx
\end{split}
\vspace{-0.0pt}
\end{equation}
%Introducing the pseudo-state $x_\rho$ leads to a homotopy-like performance metric for the system. This metric covers all parameter perturbations and higher-order errors from the nominal LTV performance to the maximum bounds. In other words, the influence of the parameter variation vanishes for $x_\rho=0$ and is completely recovered for $x_\rho=1$. 
Recall that the parameter perturbations $\rho_\Delta$ are confined to lie in a compact set $\mathcal{P}$.
% this out
%In general, the Taylor series expansion's higher-order-terms are small in the proximity of the reference trajectory.
%Thus, for each point in%Other methods include, for example. sampling to provide a bound.  time bounds on $\epsilon_f$ and $\epsilon_h$ can be determined. 
%This effectively confines them to sets $\mathcal{E}_f$ and $\mathcal{E}_h$.
%Reference \cite{Takarics2015} provides a formal bound on $\epsilon_f$ and  $\epsilon_h$  derived from the Lipschitz constant of the nonlinear system. 
% this out
Confining the term $\rho_\Delta(t)$ %$\epsilon_f$, and $\epsilon_h$
to its set allows us to treat it as an uncertainty $\Delta$ providing an uncertain LTV system given by
the known state space model $\hat{N}$:
%%\begin{equation}
%%\begin{split}
%%\label{eq:uLTVext}
%%\bmtx \dot{x}_\Delta(t)\\ 0 \\v(t)\\ e_\Delta(t)\\ y_\Delta (t)\emtx \!\!&=\!\!
%%\bmtx A(t) \!\!&\!\! 0 \!\!&\!\! E(t) \!\!&\!\! B_d(t) \!\!&\!\! B_u(t)\\
%%0  \!\!&\!\! 0 \!\!&\!\! 0 \!\!&\!\! 0 \!\!&\!\! 0\\
%%0 \!\!&\!\! 1\!\! &\!\! 0 \!\!&\!\! 0\!\! &\!\! 0\\
%%C_{e}(t)\!\!&\!\! 0 \!\!&\!\! F_h(t) \!\!&\!\! D_{ed}(t) \!\!&\!\! D_{eu}(t)\\
%%C_{y}(t)\!\!&\!\! 0 \!\!&\!\! F_g(t) \!\!&\!\! D_{yd}(t) \!\!&\!\! D_{yu}(t)
%%\!\!\emtx \bmtx x_\Delta(t)\\ x_\rho \\ w(t)\\ d_\Delta(t) \\ u_\Delta(t)\emtx\\ \\[-17pt]
%%w&=\text{diag}(\rho_\Delta) v.
%%\end{split}
%%\end{equation}
\begin{equation}
\begin{split}
\label{eq:uLTVext}
\bsmtx \dot{x}_\Delta(t)\\ 0 \\v(t)\\ e_\Delta(t)\\ y_\Delta (t)\esmtx \!\!&=\!\!
\bsmtx A(t) & 0 & E(t) & B_d(t) & B_u(t)\\
0  & 0 & 0 & 0 & 0\\
0 & 1 & 0 & 0 & 0\\
C_{e}(t)& 0 & F_h(t) & D_{ed}(t) & D_{eu}(t)\\
C_{y}(t)& 0 & F_g(t) & D_{yd}(t) & D_{yu}(t)
\!\!\esmtx \bsmtx x_\Delta(t)\\ x_\rho \\ w(t)\\ d_\Delta(t) \\ u_\Delta(t)\esmtx\\ \\[-14pt]
w&=\text{diag}(\rho_\Delta) v.
\end{split}
\end{equation}
%% \begin{equation}
%%\begin{split}
%%\label{eq:uLTVext}
%%\bmtx \dot{x}_\Delta(t)\\ 0 \\v(t)\\ e_\Delta(t)\\ y_\Delta (t)\emtx &=
%%\bmtx A(t) & 0 & B_w(t) & B_d(t) & B_u(t)\\
%%0 & 0 & 0 & 0 & 0\\
%%C_v(t) & E(t) & D_{vw}(t) & D_{vd}(t) & D_{vu}(t)\\
%%C_{e}(t)& 0 & F_h(t) & D_{ed}(t) & D_{ed}(t)\\
%%C_{y}(t)& 0 & F_g(t) & D_{yd}(t) & D_{yd}(t)
%%\emtx \bmtx x_\Delta(t)\\ x_\rho \\ w(t)\\ d_\Delta(t) \\ u_\Delta(t)\emtx\\ \\
%%w&=\text{diag}(\rho_\Delta) v.
%%\end{split}
%%\end{equation}
%with $\Delta := \rho_\Delta(t)$.
%The result is an uncertain LTV system. Thus, the system $\hat{H}$ can be written in LFT form, i.e., $\mathcal{F}_u(N, \Delta)$. This recovers the formulation in section \ref{sec:pre} with the certain part given by \eqref{eq:uLTV} (with the extended state vector $\hat{x}_\Delta= [x_\Delta^T, x_\rho^T]^T$) and the uncertainty $\Delta$ representing $\rho_\Delta$. % $\epsilon_f$, and $\epsilon_h$.
For a given controller, the performance of the closed loop system $\mathcal{F}_u(\mathcal{F}_L(\hat{N}, K), \Delta)$ can be quantified by \eqref{eq:E2PWC}. Recall that the pseudo-state $x_\rho$ requires a non-zero initial condition to take effect, but ``actual'' state variables $x_\Delta$ have zero initial conditions.
$\mathcal{F}_u(\mathcal{F}_L(\hat{N}, K), \Delta)$ together with the nominal output provides an upper bound for the robust performance of the nonlinear closed loop system $F_l(\tilde{N}, K)$.
Moreover, it allows to state a synthesis problem calculating an LTV controller minimizing the upper bound on the worst-case closed loop performance:
\vspace{-5.0pt}
\begin{align}
  \label{eq:robsyn}
  \inf_{K} \sup_{\Delta \in \mathcal{P}} 
  \|\mathcal{F}_u(\mathcal{F}_l(\hat{N},K),\Delta)\|_{2[0,T]}
\end{align}
%\begin{align}
%  \label{eq:robBnd1}
%  \|\mathcal{F}_u(\mathcal{F}_l(N,K),\Delta)\|_{[0,T]} < \gamma,
%  %\,\,\, \forall \Delta \in \mathcal{S}
%\end{align}
%!maybe better as the optimization!
%where $\hat{H}$ is written in LFT form and $\Delta$ contains the real parametric uncertainty representing $\rho_\Delta$.%, $\epsilon_f$, and $\epsilon_h$.
This synthesis problem can be solved as a robust LTV synthesis following the steps in \ref{ss:RobSyn}.
However, the synthesis of a robust LTV controller for a nonlinear system along a perturbed trajectory requires
partially non-zero initial conditions in the nominal synthesis and robust performance step, respectively.
\vspace{-2.5pt}
\subsection{Computational Approach}\label{ss:Algo}
The pseudo-code in Algorithm 1 describes the synthesis of an LTV controller for a nonlinear system along an uncertain trajectory.
\vspace{-3.0pt}
\begin{algorithm}[ht!]
\caption{Uncertain Trajectory Synthesis}
\label{Alg:Log-L-SHADE}
\begin{algorithmic}[1]

\State {\textbf{Input:} $\hat{N}$, $i_\text{max}$, $M_v^{(0)}$, $M_v^{(0)}$,  $\gamma_\text{IQC}^{(0)}$}
\State{\textbf{Output:} $\gamma_\text{IQC}^{(i)}$, $K^{(i)}$}
%\State{\textbf{Initialize:} $i_\text{max}$, $M_v^{(0)}$, $M_v^{(0)}$,  $\gamma_\text{IQC}^{(0)}$}
\For{$i = 1 : i_\text{max}$}
\State\parbox[t]{0.9\linewidth}{ {\textbf{Step 1:} Build scaled plant $G_\text{s}$ (\figref{fig:ScaledPlant}) from $\hat{N}$, $M_w^{(i-1)}$, $M_v^{(i-1)}$, and  $\gamma_\text{IQC}^{(i-1)}$}}
\State\parbox[t]{0.9\linewidth}{ {\textbf{Step 2:} Synthesize $K^{(i)}$ for $G_\text{s}$ with uncertain initial pseudo state ${x_\rho}$ and $\rho_\Delta$ set to $b_\rho$}}
\State\parbox[t]{0.9\linewidth}{{\textbf{Step 3:} LTV IQC analysis of closed loop with uncertain initial pseudo state $x_\rho$ and uncertain $\rho_\Delta$ using $K^{(i)}$} yielding robust performance $\gamma_\text{IQC}^{(i)}$, $M_w^{(i-1)}$, and $M_v^{(i-1)}$}
\EndFor

\end{algorithmic}

\end{algorithm}

%%%%% rewrite %%%%
\vspace{-3.0pt}
A user-defined amount of iterations $i_\text{max}$ is provided to the algorithm together with
the system $\hat{N}$ \eqref{eq:uLTVext}, with the additional input $u$ and output $y$. The uncertainty covering the parameter variation $\Delta:=\text{diag}(\rho_\Delta)$ is described by $\mathcal{S}$ with its input/output behavior bounded by the IQC in Example \ref{ex1}. Its norm bound is denoted $b_\rho$. Initial values of the scalings $M_{v}^{(0)}:=I_{n_v}$, $M_{w}^{(0)}:=I_{n_w}$, 
and $\gamma_\text{IQC}^{(0)}:= 1$ are also provided.

Each iteration starts with the construction of the scaled plant $N_\text{s}$, as described in section \ref{ss:RobSyn}.
For the first iteration $N_\text{s} = \hat{N}$ given the initialization choices.

The next step is a nominal finite horizon synthesis for the scaled plant $G_\text{s}$ under non-zero initial conditions. Here, the parameter variation $\rho_\Delta$ is treated as constant vector, with its entries chosen to the corresponding values of the uncertainty norm-bond $b_\rho$.
%Note that the synthesis accounts for the driving term, but does not cover any time-variation in the parameter perturbation.
Reference \cite{khargonekar1991h_} provides
necessary and sufficient conditions for the existence of a
$\gamma$-suboptimal controller for non-zero initial conditions. Theorem \ref{thm2} below is a special case of the results in \cite{khargonekar1991h_}.
%\textcolor{blue}{FB: do you think we need the theorem? I mean especially this one doesn't add much... it is basically just
%the one from [5] sin terminal penalty and in our notation (additional matrices, missing atm). However, a paper without theorems probably has a hard time to get through the review. If we add it I think I need to define the nonlinear system already with $u$ and $y$... otherwise it becomes untraceable}
\begin{mytheo} 
  \label{thm2}	 
  Consider an LTV system $\hat{N}$ \eqref{eq:uLTVext} with a fixed value of $\rho_\Delta$, and  $\gamma > 0$ given. Let $Q$ be a given positive definite diagonal matrix. Let $B$, $\hat{C}$, $\bar{R}$ and $\hat{R}$ be defined as follows.
  \begin{align*}
    \begin{split}
      \hat{A}&:= \bsmtx A \,\,&\,\, 0 \\ 0 \,\,&\,\, 0\esmtx \hspace{0.2in} \hat{B} := \bsmtx Eb_\rho & B_d  \esmtx  \\
      B&:= \bsmtx Eb_\rho & B_d & B_u\esmtx \hspace{0.2in} \bar{R}:=\text{diag}\{-\gamma^{2}I_{n_w+n_d}, I_{n_u}\}\\
      \hat{C}&:= \bsmtx 0 & C_e^T & C_y^T \\ 1^T & 0 & 0\esmtx^T \hspace{0.05in} \hat{R}:=\text{diag}\{-\gamma^{2}I_{n_v+n_e}, I_{n_y}\}
    \end{split}
  \end{align*}
  \begin{enumerate}		
  \item There exists an admissible output feedback controller $K$ such that $\|\mathcal{F}_l(\hat{N},K)\|_{2[0,T]}<\gamma$ if and only if the following three conditions hold:
    \begin{enumerate}			
    \item There exists a differentiable function
      $X:[0,T] \rightarrow \Sm^n$ such that $X(T) = 0$,	
      \begin{align*}
        \dot{X} + \hat{A}^{T}X + X\hat{A} - XB\bar{R}^{-1}B^TX + \hat{C}^T\hat{C} = 0
      \end{align*}
	and $X(0) < \gamma^2 Q$.
    \item There exists a differentiable function
      $Y:[0,T] \rightarrow \Sm^n$ such that $Y(0) = Q^{-1}$,		
      \begin{align*}
        -\dot{Y} + \hat{A}Y + Y\hat{A}^{T} - Y\hat{C}^T\hat{R}^{-1}\hat{C}Y + \hat{B}\hat{B}^T= 0
      \end{align*}
    \item $X(t)$ and $Y(t)$ satisfy the following point-wise in time spectral radius condition,
      \begin{equation}
        \label{ccond}
        \rho(X(t)Y(t))<\gamma^2 \text{, } \forall t \in [0,T]
      \end{equation}
    \end{enumerate}
  \item If the conditions above are met, then the closed loop performance
    $\|\mathcal{F}_l(\hat{N},K)\|_{[0,T]}<\gamma$ is achieved by the
    central controller constructed from $X$ and $Y$.
    % XXX PJS - Cut
    % \begin{align*}
    %   \dot{\hat{x}}(t) &= A_K(t)\, \hat{x}(t) + B_K(t)\, y(t)\\
    %   u(t) &= C_K(t)\, \hat{x}(t)
    % \end{align*}
    % where
    % \begin{align*}
    %   Z &  := (I - \gamma^{-2}YX)^{-1} \\
    %   A_K& := A + \gamma^{-2}B_dB_d^TX - ZYC_y^TC_y - B_uB_u^TX\\
    %   B_K& := ZYC_y^T\\
    %   C_K& :=-B_u^TX 
    % \end{align*}
  \end{enumerate}
\end{mytheo}
The reader is referred to \cite{khargonekar1991h_} for the proof and technical assumptions on the structure of system matrices.

For a given value of $\gamma>0$ and $Q$, the RDEs associated with $X$ and $Y$
are integrated backwards and forward in time, respectively. The diagonal entry in $Q$ corresponding to $x_\rho$
is chosen to one while all other entries are set to $10^6$. This choice of $R$ renders system states $x_\Delta$ approximately zero and confines $x_\rho$ to the set $\mathcal{U}:= [-1, 1]$.
If a solution to both RDEs exist and the condition $X(0) < \gamma^2 Q$ is fulfilled then the spectral radius coupling
condition \eqref{ccond} is evaluated.  
The controller achieves the closed-loop
performance $\gamma$ if all conditions are fulfilled.
A bisection yields the corresponding controller $K^{(i)}$.

% mathish
%Subsequently, the algorithm performs a robust performance analysis of the uncertain closed loop $\mathcal{F}_u(\mathcal{F}_l(N, K), \Delta)$.
%Note that $N$ and $\Delta$ originate from the decomposition of the uncertain plant $\hat{H}$ which contains the real time-varying uncertainty $\rho_\Delta$. %, $\epsilon_f$, and $\epsilon_h$.
%wordish
Subsequently, the algorithm performs a robust performance analysis of the uncertain closed loop built from $N:=\mathcal{F}_l(\hat{N}, K^{(i)})$ and $\rho_\Delta$.
%A dissipation inequality can be stated to upper bound the worst-case induced $L_2[0,T]$ gain of the interconnection $F_u(N, \Delta)$.
The next Theorem \ref{thm3} states a sufficient DLMI condition to upper
bound the worst-case performance of an uncertain system $\mathcal{F}_u(N,\Delta)$ with non-zero initial conditions. %\textcolor{blue}{FB: should be $N$ as the system with pseudo-states $\tilde{H}$ can be written as LFT leading to $N$ and $\Delta$}
%\textcolor{blue}{basically the same question as above. If it remains I need to update the matrices}
\begin{mytheo} 
  \label{thm3}
  Consider an LTV system $N$ given by \eqref{eq:uLTV} and let
  $\Delta:\mathcal{L}^{n_v}_2[0,T]\rightarrow\mathcal{L}^{n_w}_2[0,T]$
  satisfy the IQC defined by $M:[0,T] \rightarrow
  \Sm^{(n_v+n_w)}$. Assume $\mathcal{F}_u(N,\Delta)$ is
  well-posed. Let $Q : [0,T] \rightarrow \Sm^{n_{\hat{x}}}$,
  $S : [0,T] \rightarrow \R^{n_{\hat{x}} \times (n_w + n_d)}$, and
  $R : [0,T] \rightarrow \Sm^{(n_w + n_d)}$ be defined as follows.
\vspace*{-5pt}  
\begin{align*}
    D_v:&=\bsmtx D_{vw} & D_{vd} \esmtx \hspace{0.1in} D_e:=\bsmtx D_{ew} & D_{ed} \esmtx \hspace{0.1in}
    \hat{Q} := C_e^T C_e, \hspace{0.1in}  \\
        S &:= C_e^T D_e, \hspace{0.1in} 
    R  := D_e^T D_e - \gamma^2 \mbox{diag}\{0_{n_w}, I_{n_d}\} 
  \end{align*}
  If there exists $\epsilon > 0$, $\gamma>0$ and a differentiable function $P:[0,T]\rightarrow\Sm^{n_x}$ such that $P(T)\geq 0$, $P(0) <\gamma^2 Q$  and,
 \begin{align}
  \label{LMICond}
  \!\bsmtx
    \!\dot{P}\!+\!A^TP\!+\!PA & PB\\
    \!B^TP\! &\! 0\! 
  \esmtx \!+\! \bsmtx
    \!\hat{Q}\! & S\!\\
    \!S^T\! & R\!
  \esmtx
  + \bsmtx C_v^T\\ D_v^T \esmtx M \bsmtx
  C_v^T \\ D_v^T
	\esmtx^T \leq -\epsilon I
\end{align}
  then $\|\mathcal{F}_u(N,\Delta)\|_{2[0,T]} < \gamma$.
\end{mytheo}
\begin{proof}
Theorem is a corollary of Theorem 6  in \cite{Seiler2019}. The
proof provided in \cite{Seiler2019}, which combines a standard dissipation
argument \cite{willems1972dissipative1,willems1972dissipative2,hill1980dissipative}
and IQCs, requires only minor modifications.
\end{proof}
The IQC analysis step uses the algorithm proposed in \cite{Seiler2019} with an extension to uncertain initial conditions for the upper bound calculation. The same value for $Q$ as in the
synthesis is chosen, with the same implications.
%It combines the DLMI formulation in the theorem above with a related Riccati Differential Equation. 
This algorithm returns the robust performance upper bound
$\gamma_\text{IQC}^{(i)}$, and $M_v^{(i)}$ and $M_w^{(i)}$, which are used to build the scaled plant as described in section \ref{ss:RobSyn}. The algorithm commences with the next iteration and stops after a defined amount of iterations $i_\text{max}$.

\vspace{-0.5pt}
\section{Numerical Example: Space Launcher}\label{sec:NumEx}
\vspace{-3.5pt}
%\begin{itemize}
%\item launcher asu paper
%\item gravity turn für die trajectory is die ref trajectory kurz beschrieben
%\item und hier habe ich wind null 
%\item jetzt störung wind ist der wird rho und das nennen wir rhow
%\item dafür regler
%\item nur pitch regler ... der brauch x nicht gleichungen entlang der trajectory
%\item schreibe N als funktion von w, zdot, t  rest ist funktion zeit
%\end{itemize}

The numerical example concerns a robust pitch tracking control design for the rigid body motion of a Vega-like launch vehicle \cite{Biertuempfel2024}. 
During the atmospheric ascent, the space launcher performs a gravity turn maneuver by tracking a pre-calculated pitch program. 
The gravity turn is the reference trajectory $\mathcal{T}$ and assumes zero nominal wind, i.e., $w_\mathcal{T} = 0$
%The maneuver minimizes loads and maximizes forward acceleration by guaranteeing zero lateral acceleration
%for the nominal wind $w_\mathcal{T} = 0$. 
However, the launch performance is highly sensitive wind perturbations $w_\Delta$. Their influence must be mitigated by
the controller. 

The launcher's nonlinear equations of motion are formulated with respect to its instantaneous center of gravity (CoG) $G$ in a body-reference coordinate system and given by: %The $x_b$-axis is aligned with the launcher's symmetry axis and is defined positive in the direction of forward travel. The $z_b$-axis points downward, forming a right-hand system with the $y_b$-axis. 
%The nonlinear EoM are given by: 
\begin{equation}
\label{eq:EoMNonLin}
\begin{split}
\ddot{\theta}(t) =& \frac{N(w, \dot{x}, \dot{z}, t) l_{GP}(t)}{J(t)} - \dot{\theta}\frac{\dot{J}(t)}{J(t)} \\ 
&- \frac{T(t) l_{CG}(t)}{J(t)}\sin{\delta_{\text{TVC}}(t)} \\
\ddot{x}(t) =&  \frac{T(t)\cos{\delta_\text{TVC}(t)}-A(w, \dot{x}, \dot{z}, t)}{m(t)} \\
&- g_0 \sin{\theta}(t) - \dot{\theta}(t)\dot{z}(t)\\
\ddot{z}(t) =& -\frac{N(w, \dot{x}, \dot{z}, t)}{m(t)} - \frac{T(t)}{m(t)}\sin{\delta_{\text{TVC}}(t)} \\
&+ g_0 \cos{\theta(t)} - \dot{\theta}(t)\dot{x}(t)\\
\end{split}
\end{equation}
%Note that state $\dot{x}_b$ has been removed from the linear model following common practice in launcher pitch control design, see, e.g.,~\cite{Greensite1967b}.
The variable $\theta$ denotes the pitch angle, while the axial and normal accelerations are denoted $\ddot{x}$ and $\ddot{z}$.
The variables $A$ and $N$ are the axial and normal aerodynamic acting on a time-dependent reference point $P$. Both forces depend nonlinearly on the wind velocity $w$ (aligned with the $z_b$-axis), the forward velocity $\dot{x}$, $\dot{z}$, and time $t$.  
The launcher's time-dependent thrust $T$ acts at the reference point $C$. It can be rotated by the angle $\delta_{\text{TVC}}$ using a thrust vector control (TVC) system. 
The overall moment of inertia $J$, mass $m$ and CoG depend on time due to the thrust profile. The variables $l$ denote the distance between reference points given in the subscript. Gravitational acceleration is denoted by $g_0$.
%\begin{figure}[!b]
%\centering
%\def\svgwidth{1\columnwidth}
%\import{figures/}{Launcher_Vega_CEAS_updt.pdf_tex}
%\caption{Expandable launch vehicle in body fixed reference frame}
%\label{fig:Launcher}
%\end{figure}

While the launcher parameters are assumed known, 
the wind variation $w_\Delta$ along the reference trajectory $\mathcal{T}$ is assumed $\pm 20\,$m/s uncertain. Hence, the wind variation is treated as an exogenous parameter $\rho$. 
Linearizing (\ref{eq:EoMNonLin}) about the gravity turn $\mathcal{T}$ results in the LTV system: % nur explizit das E angeben
%%\begin{equation}% no uncertainty or pseudo state
%%  \label{eq:LTVpitch}
%%  \begin{split}
%%\!\bmtx \dot{\theta}_b \\ \ddot{\theta}_b \\ \ddot{z}_b \\ \dot{w}_z\emtx \!\!\!\!&=\!\!\! 
%%\bsmtx 0 & 1 & 0  & 0\!\\ 0 & -\frac{\dot{J}_{yy_0}}{J_{yy_0}} & \frac{l_{GP}}{J_{yy}\dot{x}_{b_0}}\frac{\partial N}{\partial \alpha}|_0\!\!\\ -g_0\sin{\theta_{b_0}} &  -\dot{x}_{b_0} & - \frac{1}{m\dot{x}_{b_0}} \frac{\partial N}{\partial \alpha}|_0 & \frac{l_{GA}}{J_{yy}\dot{x}_{b_0}} \frac{\partial N}{\partial \alpha}|_0\!\!\\
%%    -g_0\cos{\theta_{b_0}} &  0 &   -\dot{\theta}_{b_0} - \frac{1}{m_0\dot{x}_{b_0}} \frac{\partial N}{\partial \alpha}|_0 &  \frac{1}{m_0(t)\dot{x}_{b_0}}\frac{\partial N}{\partial \alpha}|_0\!\! \\ 0 & 0 & 0 & 0\!\!  \esmtx  \!\!\!\! \bmtx \theta_b \\ \dot{\theta}_b \\ \dot{z}_b \\ w_z\emtx
%%    \\&+ \bsmtx 0 & 0\\
%%     -\frac{T_0 l_{CG_0}}{J_{yy_0}} \\
%%     -\frac{T_0}{J_{yy_0}} \esmtx
%%      \delta_{\text{TVC}},
%%      \end{split}
%%\end{equation}
\begin{equation}
  \label{eq:LTVpitch}
  \begin{split}
\!\!\!\!\bsmtx \!\!\dot{\theta}_\Delta\!\! \\ \!\!\ddot{\theta}_\Delta\!\! \\ \!\!\ddot{z}_\Delta\!\! \\ \!\!0\!\!\esmtx \!\!\!&=\!\!\! 
\bsmtx 0 & 1 & 0  & 0\!\\ 
0 & -\frac{\dot{J}_{\mathcal{T}}}{J_{\mathcal{T}}} & \frac{l_{GP_\mathcal{T}}}{J_{\mathcal{T}}\dot{x}_{\mathcal{T}}}\frac{\partial N}{\partial \dot{z}}|_\mathcal{T}\!\! & \frac{l_{GP_\mathcal{T}}}{J_{\mathcal{T}}\dot{x}_{\mathcal{T}}}\frac{\partial N}{\partial w}|_\mathcal{T} w_\Delta \\ 
-g_0\sin{\theta_{\mathcal{T}}} &  -\dot{x}_{\mathcal{T}} & - \frac{1}{m_{\mathcal{T}}\dot{x}_{\mathcal{T}}} \frac{\partial N}{\partial \dot{z}}|_\mathcal{T} & \frac{1}{m_{\mathcal{T}}\dot{x}_{\mathcal{T}}} \frac{\partial N}{\partial w}|_\mathcal{T} w_\Delta \!\!\\
    0 & 0 & 0 & 0\!\!  \esmtx  \!\!\!\! \bsmtx \!\theta_\Delta\!\! \\\! \dot{\theta}_\Delta\!\! \\ \!\dot{z}_\Delta\!\! \\ \!x_\rho\!\! \esmtx
    \\&+ \bsmtx 0 \\
     -\frac{T_\mathcal{T} l_{CG_\mathcal{T}}}{J_{\mathcal{T}}} \\
     -\frac{T_\mathcal{T}}{J_{\mathcal{T}}} \esmtx
      \delta_{\text{TVC}_\Delta}.
      \end{split}
\vspace{-10pt}
\end{equation}
The state $\dot{x}_b$ has been truncated from the linear model following common practice in launcher pitch control design yielding a typical system order for launcher control designs, see, e.g.,~\cite{Greensite1967b}. All entries in \eqref{eq:LTVpitch} are time-varying. The coefficients in \eqref{eq:LTVpitch} are calculated via numerical linearization for a time segment from $25\,$s to $80\,$s after lift-off, with a grid density of $0.25\,$s.

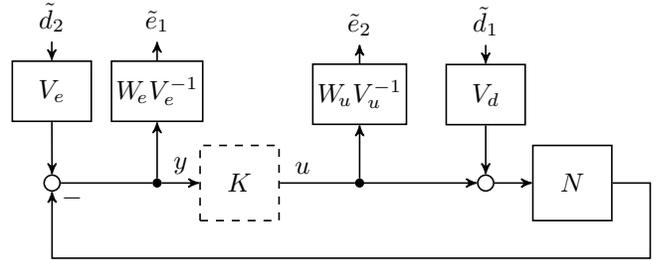
\begin{figure}[ht!] 
\vspace*{-12pt}
	\centering\usetikzlibrary{positioning,plotmarks, matrix, arrows, calc, shapes}
\tikzstyle{blockdiag}	= [node distance=5mm, >=stealth', semithick]
\tikzstyle{block}			= [draw, rectangle, minimum width=1.05cm, minimum 
height=.8cm]
\tikzstyle{sum} = [draw,circle,inner sep=0pt, minimum size=6pt]
\tikzstyle{gain} = [draw,regular polygon, regular polygon 	sides=3,thick,minimum height=3em,minimum width=4em, rotate=30]
\tikzstyle{bguide} = [rectangle,minimum height=3em,minimum	width=4em]
\tikzstyle{line} = [thick]
\tikzstyle{branch} = [circle,inner sep=0pt,minimum size=1mm,fill=black,draw=black]
\tikzstyle{guide} = [anchor=center]

\begin{tikzpicture}[blockdiag, auto]

% Blocks
\node[block, minimum height=1.0cm] (Plant) {$N$};
\node[sum, left=of Plant, yshift=-0.0cm] (SumP) {};
\node[branch, left=of SumP, xshift=-1.0cm] (BranchP) {};
\node[block, above=of BranchP, xshift=0mm,yshift=0.21cm] (Wu) {$\!W_{\!u}V_{u}^{-1}\!$};
\node[block, above=of SumP,yshift=0.15cm] (Wd) {$\!\!V_d\!\!$};
\node[block, dashed,left=of BranchP, xshift=-0.50cm, minimum height=1.0cm] (Controller) {$K$};

\node[above=of Plant, yshift=-.75cm, minimum width=1.2cm](Pd) {};

\node[branch, left=of Controller, xshift=0cm] (BranchE) {};
\node[sum, left=of BranchE, xshift=-0.72cm] (SumE) {};
\node[block, above=of BranchE,yshift=0.25cm] (We) {$\!W_{\!e}V_{e}^{-1}\!$};
\node[block, above=of BranchE,yshift=0.25cm, xshift=-14mm] (W1) {$\!\!V_e\!\!$};

% 
% 
% \node[block, right=of Wd, xshift = 1.4cm] (Weta) {$W_z$};
% \node[block, right=of Weta] (Wy) {$W_y$};
% \node[block, above=of Sum2] (Wn) {$W_2$};

% % Conncections
%\draw[->] (Pd) -| (SumP);
%\draw[->] (Plant) -- (SumP);
\draw[<-] (SumE) -- (W1);
\draw[<-] (W1.north) -- +(-0,0.25cm) node[above]{$\tilde{d}_2$};
\draw[<-] (Wd.north) -- +(0,0.25cm) node[above]{$\tilde{d}_1$};
\draw[->] (We.north) -- +(0, +.25cm) node[above]{$\tilde{e}_1$};
\draw[->] (Wu.north) -- +(0, +.25cm) node[above]{$\tilde{e}_2$};

\draw[-] (SumE) -- (BranchE);
\draw[->] (BranchE) -- (Controller) node[pos=0.5] {$y$};
\draw[->] (BranchE) -- (We);
\draw[-] (Controller) -- (BranchP) node[pos=0.3] {$u$};
\draw[->] (Wd) -- (SumP.north);
\draw[->] (BranchP) -- (Wu);
\draw[->] (BranchP) -- (SumP);
\draw[->] (SumP) -- (Plant);

 \draw[->] ($(Plant.east)+(0,-0.0cm)$) -| +(+.5cm,-1.0cm) -| (SumE.south) node[pos=0.95,swap] {$-$};
% \draw[->] ($(Plant.south east)!.3!(Plant.north east)$) +(+.8cm,-0.0cm) node[branch] {} -| (Wy);
% \draw[<-] (Wn.north) -- +(0, +.5cm)node[left, name=d2]{$w_1$};
% \draw[<-] (Wd.north) -- +(0, +.5cm)node[left, name=d1]{$w_2$};
% 
% \draw[->] (Wn.south) -- (Sum2.north);
% \draw[->] (Wd.south) -- (Sum1.north);
% \draw[<-] (Wu.south) -- (Con1.north);
% \draw[->] (Sum2.east) node[below right]{} -- (Controller.west);
% \draw[->] (Controller.east) -- (Sum1.west) node[pos=0.2]{$u$};
% \draw[->] (Sum1.east) -- (Plant.west);

;\end{tikzpicture} 
	\caption{Uncertain output feedback interconnection}
	\label{fig:StructSynth} %at 3:48am seismic activity was registered in Hamburg, as Julian woke up screaming in the moment I changed the signal names
\vspace{-6.0pt}
\end{figure}

The robust control design requires a weighting structure to enforce meaningful closed loop requirements, such as tracking and limited control authority.
A suitable (mixed-sensitivity) weighting structure for a space launcher is provided in \cite{Biertuempfel2024} including tuning guidelines. The weighting scheme is depicted in Fig.~\ref{fig:StructSynth}.
% Fig. weighted
%\begin{figure}[!b] 
%	\centering\input{figures/control_design_MixedSensitivity_sdf.tikz}
%	\caption{Weighted four-block mixed sensitivity problem.}
%	\label{fig:StructSynth}
%\end{figure}
%%%%%%%%%%%%%%%%%%%%%%%%%%%%% if we want insert the explicit norm problem again %%%%%%%%%%%%%%%%%%%%%
%It represents the weighted closed loop
%\begin{equation}\label{eq:control_design_MixedSensitivity_weighted} 
%\begin{bmatrix}
%z_1 \\ z_2
%\end{bmatrix}
%\!\!=\!\!
%\begin{bmatrix}
%\!W_{e}V_e^{-1}\!\!\!\! & 0\\0 & \!\!\!\!W_{u}V_u^{-1}\!\!
%\end{bmatrix}
%\!\!\!
%\begin{bmatrix} 
%	-SP\!\!\!\! &\!\! S     \\
%	-KSP \!\!\!\!& \!\!KS    
%\end{bmatrix}
%\!\!\!
%\begin{bmatrix}
%V_d \!\!\!\!&\!\! 0 \\ 0 \!\!\!\!&\!\! V_e
%\end{bmatrix}
%\!\!\!
%\begin{bmatrix}
%w_1 \\ w_2
%\end{bmatrix}\!\!.
%\end{equation}

The pitch angle $\theta$ and the pitch rate $\dot{\theta}$ are used as feedback signals. 
The weight $W_\text{e}$ imposes tracking and disturbance rejection requirements. Here, $\theta$ shall be tracked up to a desired closed loop bandwidth of $0.75\,\text{rad}/\text{s}$. Thus, $W_e$ is selected with integral behavior in the $\theta$ channel up to a frequency of $0.75\,\text{rad/s}$ and a magnitude of $0.5$ for robustness. A constant weight of $0.5$ is selected in the $\dot{\theta}$ channel following the same reasoning.
The weight $W_u$ imposes control limitations and noise rejection requirements. To limit TVC activity to frequencies less than $10\,\text{rad/s}$, $W_u$ is selected with unit magnitude up to $25\,\text{rad/s}$ and differentiating behavior above $25$\,rad/s.
The static weight $V_e$ balances the two output errors. It is chosen as $1.0\,\text{deg}$ for $\theta$, $2.5\,\text{deg/s}$ for $\dot{\theta}$. 
$V_e$ specifies the available control effort against the chosen errors and is selected as $5\,\text{deg}$.

The controller is synthesized using Algorithm 1 with $N_\text{syn} = 4$ iterations.
An updated version of the algorithm in \cite{Seiler2019} is used to perform the IQC analysis step with a tolerance of $5 \cdot 10^{-5}$, five iterations between the LMI and RDE solution, and $25$ and $10$ evenly spaced LMI and spline grid points on the considered time segment, respectively. The algorithm converged to a $\gamma$-value of $11.34$.
The synthesis required $1\,$h and $45\,$min on a standard desktop computer with an Intel-Core i9 processor and $16\,$GB memory. 

For comparison a nominal controller with non-zero initial conditions was synthesized following Theorem \ref{thm2}. The synthesis accounts for $x_\rho$. Instead of treating $w_\Delta$ as time-varying uncertainty, it is assumed with a fixed value of $20\,$m/s. Thus, the design only accounts for an (unknown) additional constant wind disturbance.
The same weights $W$ and $V$ are used as in the robust synthesis. %Although, this controller considers the pseudo state it cannot account for time variations in the wind disturbance.

Simulations are conducted in Matlab Simulink of the closed loops formed by nonlinear launcher dynamics and the nominal and the robust controller, respectively. 
Wind disturbance is injected into the system. It is constructed by superimposing the wind profile from Vega flight VV05 \cite{Simplicio2016} with $1000$ unique cases of severe Dryden turbulence \cite{Hoblit1988}. Launcher certification processes commonly use turbulence models to account for uncertainty in wind estimations. Fig. \ref{fig:VV05turb} shows one of the injected wind profiles.
Both controllers show no instability for these wind profiles.
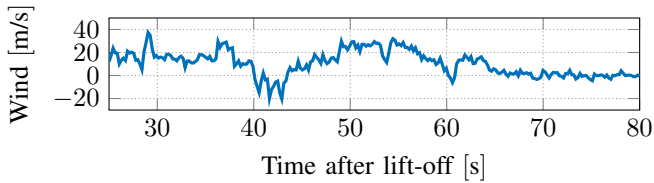
\begin{figure}[b!]
\vspace{-19pt}
\centering
\begin{tikzpicture}
\definecolor{blue1}{RGB}{222,235,247}
\definecolor{blue2}{RGB}{158,202,225}
\definecolor{blue3}{RGB}{49,130,189}
\begin{axis}[ width = 1.0\columnwidth, height = 0.325\columnwidth, % Grösse
  	grid=major, 
   grid style={densely dotted,white!60!black}, 
   xlabel= Time after lift-off ${[\text{s}]}$, 		% x label
   ylabel=  Wind ${[\text{m/s}]}$, 	% y label
   legend style={at={(0.5,0.01)},anchor=south east},
   legend cell align = {left},
   xmin = 25, xmax = 80, ymin = -30, ymax = 50, % Hier kannst du die Achsenabschnitte definieren 
        ]
% DI        
%\addplot[blue, line width = 2, mark = o,  mark size =4pt] table[x expr = \thisrowno{0} ,y expr = \thisrowno{4} ,col sep=comma] {figures/E2P.csv};
%\addlegendentry{b = 1e-6} % Legendeneintrag

%\addplot[name path = A, gray!30, line width = 1,  no marks, forget plot] table[x expr = \thisrowno{0} ,y expr = \thisrowno{2} ,col sep=comma] {figures/WindData.csv};
%\addplot[name path = B,gray!30, line width = 1,  no marks] table[x expr = \thisrowno{0} ,y expr = \thisrowno{3} ,col sep=comma] {figures/WindData.csv};\label{Envelope}
%\addplot [gray!30, forget plot] fill between [of = A and B];

%\addplot[black!60, line width = 0.8,  no marks] table[x expr = \thisrowno{0} ,y expr = \thisrowno{5} ,col sep=comma] {figures/WindData.csv};\label{pl:Exmp}
%\addplot[black!60, line width = 0.8,  no marks, forget plot] table[x expr = \thisrowno{0} ,y expr = \thisrowno{7} ,col sep=comma] {figures/WindData.csv};\label{pl:Exmp}

%\addplot[red!80, line width = 1, dotted,  no marks] table[x expr = \thisrowno{0} ,y expr = \thisrowno{4} ,col sep=comma] {figures/WindData.csv};\label{pl:mean}
\addplot[RoyalBlue, line width = 1.2,  no marks] table[x expr = \thisrowno{0} ,y expr = \thisrowno{1} ,col sep=comma] {figures/WindNew.csv};\label{TurbLongPlot}
%\addplot[ForestGreen, line width = 0.75,  no marks] table[x expr = \thisrowno{0} ,y expr = \thisrowno{2} ,col sep=comma] {figures/WindNew.csv};\label{TurbLongPlot}

%\addlegendentry{Envelope Wind Filter}
%\addlegendentry{Selected Wind Signal}
%\addlegendentry{Steady Wind Profile}
%\addlegendentry{VV05 Estimate}

%\addplot[black, line width = 1,  no marks] table[x expr = \thisrowno{7} ,y expr = \thisrowno{16} ,col sep=comma] {figures/WandT2QalphaPlot.csv};\label{QalphaSignal5}

%\addplot[black, line width = 1,  no marks] table[x expr = \thisrowno{8} ,y expr = \thisrowno{17} ,col sep=comma] {figures/WandT2QalphaPlot.csv};\label{QalphaSignal6}

%\addplot[red, line width = 0.5,  no marks] table[x expr = \thisrowno{1} ,y expr = \thisrowno{3} ,col sep=comma] {figures/E2PFuckItExt.csv};\label{DI:b0.09}

%\addlegendentry{b = 0.09} % Legendeneintrag

%\addlegendentry{b = 0.11} % Legendeneintrag

%\draw[-, black, dashed, line width = 2] (axis cs:30,220000)--(axis cs:95,220000);
%\addplot[black, loosely dashdotted, no marks, line width = 2] coordinates{(0, 220000) (95, 220000)};

%\addlegendentry{$Q\alpha_{max} = 220000Pa^\circ$} % Legendeneintrag

% LMI (aftwards as no legend entry)

\end{axis}
\end{tikzpicture}
\vspace*{-17pt}
\caption{ Representative disturbance wind signal}
\label{fig:VV05turb}
\vspace*{-5.5pt}
\end{figure}
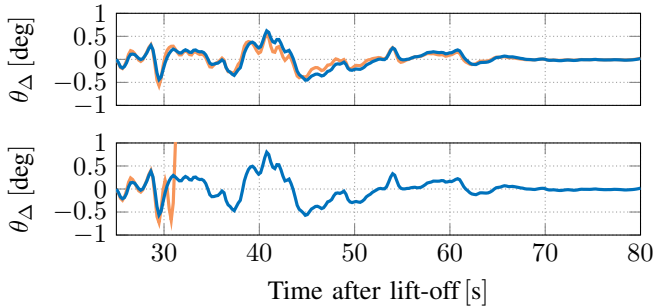
\begin{figure}[b!]
\vspace*{-7.5pt}
\centering
\begin{tikzpicture}
\definecolor{blue1}{RGB}{222,235,247}
\definecolor{blue2}{RGB}{158,202,225}
\definecolor{blue3}{RGB}{49,130,189}
%\begin{axis}[ width = 1.0\columnwidth, height = 0.5\columnwidth, % Grösse
%  	grid=major, 
%   grid style={densely dotted,white!60!black}, 
%   xlabel= Time after lift-off ${[\text{s}]}$, 		% x label
%   ylabel= Pitch Error ${[\text{deg}]}$, 	% y label
%   legend style={at={(0.5,0.01)},anchor=south east},
%   legend cell align = {left},
%   xmin = 25, xmax = 80, ymin = -0.9, ymax = 0.9, % Hier kannst du die Achsenabschnitte definieren 
%        

\begin{groupplot}[group style={
                      	group name=myplot,
                      	group size= 1 by 2,
                        vertical sep=0.5cm,
                        horizontal sep = 0.0cm},
                      	height=0.325\columnwidth,%0.33 for double column
                      	width = 0.99\columnwidth,%0.49 for double column
                      	xmajorgrids=true,
			ymajorgrids=true,
			 grid style={densely dotted,white!60!black},
			  xmin = 25, xmax = 80,
			  ymin = -1, ymax = 1,
			   ]]

\nextgroupplot[	%title=\emph{Nominal Turbulence},
				  ylabel= $\theta_\Delta\,{[\text{deg}]}$,
				 %xlabel= $\text{Time after lift-off}  \,{[\text{s}]}$,
			  	%ymin = 0, ymax = 8,
xticklabel=\empty,
				 ]
% plots with turb x1
\addplot[Peach, line width = 1.2,  no marks] table[x expr = \thisrowno{0} ,y expr = \thisrowno{1} ,col sep=comma] {figures/ResultsNew.csv};\label{pl:Nom1}
\addplot[RoyalBlue, line width = 1.2,  no marks] table[x expr = \thisrowno{0} ,y expr = \thisrowno{2} ,col sep=comma] {figures/ResultsNew.csv};\label{pl:Rob1}

\nextgroupplot[	%title=\emph{Scaled Turbulence},
				  ylabel= $\theta_\Delta\,{[\text{deg}]}$,
				 xlabel= $\text{Time after lift-off}  \,{[\text{s}]}$,
				 ]
% plots with turb x1.29
\addplot[Peach, line width = 1.2,  no marks] table[x expr = \thisrowno{0} ,y expr = \thisrowno{5} ,col sep=comma] {figures/ResultsNew.csv};\label{pl:Nom2}
\addplot[RoyalBlue, line width = 1.2] table[x expr = \thisrowno{0} ,y expr = \thisrowno{6} ,col sep=comma] {figures/ResultsNew.csv};\label{pl:Rob2}

\end{groupplot}
\end{tikzpicture}
\vspace*{-15pt}
\caption{Controller tracking performance for unscaled (top) and scaled disturbance (bottom): Robust (\ref{pl:Rob1}); Nominal  (\ref{pl:Nom1})}
\label{fig:Results}
\vspace*{-8pt}
\end{figure}
Fig.~\ref{fig:Results} depicts the resulting pitch error w.r.t. to the pitch program of the nominal (\ref{pl:Nom1}) and robust closed loop (\ref{pl:Rob1}) for the wind signal in Fig.~\ref{fig:VV05turb}.
The turbulence component is then scaled up to assess the robustness of the designs. For a scaling-factor of $1.3$, the robust controller failed to stabilize the system in $4\%$ of the cases. The nominal controller failed $13\%$; about factor three worse. The unstable cases for the robust controller could be further reduced by increasing the uncertainty level. This was not pursued in the paper to uphold a fair comparison.
%An increase by $1.3$, results in $xxx$ deviations for the nominal and only $yyy$ for the robust controller, respectively. 
Fig.~\ref{fig:Results} shows the closed loop behavior for the wind profile in Fig.~\ref{fig:VV05turb} and a scaling of $1.3$. The nominal controller becomes unstable (\ref{pl:Nom2}), while the robust controller (\ref{pl:Rob2}) can still stabilize the system and closely track the pitch profile. For this particular wind profile, a first instability occurs for a factor of 1.55 exceeding the nominal controller by $20\%$.
In general, the robust controller demonstrates significantly increased robustness against fast-varying wind disturbances.
% Thus, the robust controller provides higher robustness against fast-varying wind disturbances.

\vspace{-5pt}
\section{Conclusion}
The paper presents a novel approach for the synthesis of robust controllers for nonlinear systems along uncertain trajectories.
It formulates time-varying parameter deviations from the nominal trajectory as uncertain initial conditions of an extended state space system in conjunction with time-varying uncertainties. The latter are bounded by integral quadratic constraints. Thus, the controller can synthesized inside the LTV IQC framework. A realistic example
demonstrates the capabilities of the approach.
\vspace{-2.5pt}
%\section*{References}
\def\url#1{}
\bibliographystyle{IEEEtran}
\bibliography{ObserverBasedLTVshort}

\begin{thebibliography}{10}
\providecommand{\url}[1]{#1}
\csname url@samestyle\endcsname
\providecommand{\newblock}{\relax}
\providecommand{\bibinfo}[2]{#2}
\providecommand{\BIBentrySTDinterwordspacing}{\spaceskip=0pt\relax}
\providecommand{\BIBentryALTinterwordstretchfactor}{4}
\providecommand{\BIBentryALTinterwordspacing}{\spaceskip=\fontdimen2\font plus
\BIBentryALTinterwordstretchfactor\fontdimen3\font minus
  \fontdimen4\font\relax}
\providecommand{\BIBforeignlanguage}[2]{{%
\expandafter\ifx\csname l@#1\endcsname\relax
\typeout{** WARNING: IEEEtran.bst: No hyphenation pattern has been}%
\typeout{** loaded for the language `#1'. Using the pattern for}%
\typeout{** the default language instead.}%
\else
\language=\csname l@#1\endcsname
\fi
#2}}
\providecommand{\BIBdecl}{\relax}
\BIBdecl

\bibitem{Hosovsky2016}
A.~Ho{\v{s}}ovsk{\'{y}}, J.~Pite{\v{l}}, K.~{\v{Z}}idek,
  M.~T{\'{o}}thov{\'{a}}, J.~S{\'{a}}rosi, and L.~Cveticanin, ``Dynamic
  characterization and simulation of two-link soft robot arm with pneumatic
  muscles,'' \emph{Mech. and Mach. Theo.}, vol. 103, pp. 98--116, 2016.

\bibitem{Biertuempfel2021a}
F.~Biertümpfel, S.~Bennani, and H.~Pfifer, ``Time‐varying robustness
  analysis of launch vehicles under thrust perturbations,'' \emph{Adv. Contr.
  Appl.}, vol.~3, no.~4, 2021.

\bibitem{Capolupo2024}
F.~Capolupo and A.~Rinalducci, ``Descent and landing trajectory and guidance
  algorithms with divert capabilities for moon landing,'' in \emph{AIAA SCITECH
  2024 Forum}, 2024.

\bibitem{Limebeer1992}
D.~J.~N. Limebeer, B.~D.~O. Anderson, P.~P. Khargonekar, and M.~Green, ``A game
  theoretic approach to $\mathcal{H}^\infty$ control for time-varying
  systems,'' \emph{{SIAM} J. Control Optim.}, vol.~30, no.~2, pp. 262--283,
  1992.

\bibitem{Biertumpfel2022}
F.~Biert\"umpfel, J.~Theis, and H.~Pfifer, ``Observer-based synthesis of finite
  horizon linear time-varying controllers,'' in \emph{2022 American Contr.
  Conf. (ACC)}, vol.~30.\hskip 1em plus 0.5em minus 0.4em\relax IEEE, 2022, pp.
  2956--2961.

\bibitem{OBrien1999}
R.~O'Brien and P.~Iglesias, ``Robust controller design for linear, time-varying
  systems,'' \emph{Eur. J. Control}, vol.~5, no. 2-4, pp. 222--241, 1999.

\bibitem{Pirie2002}
C.~Pirie and G.~E. Dullerud, ``Robust controller synthesis for uncertain
  time-varying systems,'' \emph{{SIAM} J. Control Optim.}, vol.~40, no.~4, pp.
  1312--1331, 2002.

\bibitem{Farhood2008}
M.~Farhood and G.~E. Dullerud, ``Control of systems with uncertain initial
  conditions,'' \emph{IEEE Trans. Autom. Contr.}, vol.~53, no.~11, pp.
  2646--2651, 2008.

\bibitem{Buch2021}
J.~Buch and P.~Seiler, ``Finite horizon robust synthesis using integral
  quadratic constraints,'' \emph{Int. J. Rob. Nonlin. Contr.}, vol.~31, no.~8,
  pp. 3011--3035, 2021.

\bibitem{Biertuempfel2023a}
F.~Biertümpfel, J.~Theis, and H.~Pfifer, ``Robustness analysis of nonlinear
  systems along uncertain trajectories,'' \emph{IFAC-PapersOnLine}, vol.~56,
  no.~2, pp. 5831--5836, 2023.

\bibitem{Seiler2015}
P.~Seiler, ``Stability analysis with dissipation inequalities and integral
  quadratic constraints,'' \emph{{IEEE} Trans. Autom. Contr.}, vol.~60, no.~6,
  pp. 1704--1709, 2015.

\bibitem{Green1995}
M.~Green and D.~J.~N. Limebeer, \emph{Linear Robust Control}.\hskip 1em plus
  0.5em minus 0.4em\relax Upper Saddle River, NJ, USA: Prentice-Hall, Inc.,
  1995.

\bibitem{Seiler2019}
P.~Seiler, R.~M. Moore, C.~Meissen, M.~Arcak, and A.~Packard, ``Finite horizon
  robustness analysis of {LTV} systems using integral quadratic constraints,''
  \emph{Automatica}, vol. 100, pp. 135--143, 2019.

\bibitem{khargonekar1991h_}
P.~P. Khargonekar, K.~M. Nagpal, and K.~R. Poolla, ``{$H_\infty$} control with
  transients,'' \emph{SIAM J. on Control Optim.}, 1991.

\bibitem{willems1972dissipative1}
J.~C. Willems, ``Dissipative dynamical systems part i: General theory,''
  \emph{Arch. for rat. mech. and analysis}, vol.~45, no.~5, pp. 321--351, 1972.

\bibitem{willems1972dissipative2}
------, ``Dissipative dynamical systems part ii: Linear systems with quadratic
  supply rates,'' \emph{Arch. for rat. mech. and analysis}, vol.~45, no.~5, pp.
  352--393, 1972.

\bibitem{hill1980dissipative}
D.~J. Hill and P.~J. Moylan, ``Dissipative dynamical systems: Basic
  input-output and state properties,'' \emph{J. of the Franklin Inst.}, vol.
  309, no.~5, pp. 327--357, 1980.

\bibitem{Biertuempfel2024}
F.~Biertümpfel, H.~Pfifer, and J.~Theis, ``Robust space launcher control with
  time-varying objectives,'' \emph{J. Guid., Contr., Dyn.}, vol.~47, no.~5, pp.
  934--944, 2024.

\bibitem{Greensite1967b}
A.~L. Greensite, ``Analysis and design of space vehicle flight control systems.
  volume {VII} - attitude control during launch,'' NASA Marshall Space Flight
  Center; Huntsville, AL, Tech. Rep., 1967.

\bibitem{Simplicio2016}
P.~Simplicio, S.~Bennani, A.~Marcos, C.~Roux, and X.~Lefort, ``Structured
  singular-value analysis of the vega launcher in atmospheric flight,''
  \emph{J. Guid., Contr., Dyn.}, vol.~39, no.~6, pp. 1342--1355, 2016.

\bibitem{Hoblit1988}
F.~M. Hoblit, \emph{Gust Loads on Aircraft: Concepts and Applications}.\hskip
  1em plus 0.5em minus 0.4em\relax American Institute of Aeronautics and
  Astronautics, 1988.

\end{thebibliography}

\end{document}